%% file: main.tex
\documentclass{article}

\PassOptionsToPackage{numbers, compress}{natbib} 
\usepackage[final]{neurips_2025}

\usepackage[utf8]{inputenc} 
\usepackage[T1]{fontenc}    
\usepackage{hyperref}       
\usepackage{url}            
\usepackage{booktabs}       
\usepackage{amsfonts}       
\usepackage{xspace}         
\usepackage{nicefrac}       
\usepackage{microtype}      
\usepackage{xcolor}         

\usepackage{amsmath}
\usepackage{amssymb}
\usepackage{mathtools}
\usepackage{amsthm}
\usepackage{enumitem}

\usepackage{multicol}
\usepackage{longtable}
\usepackage{booktabs} 
\usepackage{makecell}
\usepackage{listings}
\usepackage{subcaption}

\usepackage[table]{xcolor}

\usepackage{wrapfig}       
\usepackage{lipsum} 

\newcommand{\reddownarrow}[1]{\textcolor{red}{$\downarrow\,#1$}}

\newcommand{\tool}{Repo2Run}

\newcommand{\TitleIconHeight}{1.2em} 
\newcommand{\TitleIconGap}{1.7em}    
\title{%
\raisebox{-0.3ex}{
  \makebox[0pt][l]{
    \includegraphics[height=\TitleIconHeight]{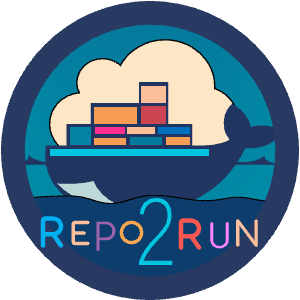}%
}}%
\hspace*{\TitleIconGap}
\tool: Automated Building Executable Environment for Code Repository at Scale
}

\author{%
  \textbf{Ruida Hu}$^{1}$\thanks{Work done during an internship at ByteDance.} \quad
  \textbf{Chao Peng}$^{2}$\thanks{Corresponding authors.} \quad
  \textbf{Xinchen Wang}$^{1}$\footnotemark[1] \quad
  \textbf{Junjielong Xu}$^{2}$ \quad
  \textbf{Cuiyun Gao}$^{1}$\footnotemark[2]
  \\ \\
  $^{1}$Harbin Institute of Technology, Shenzhen \qquad $^{2}$ByteDance, Beijing \\
  \texttt{\{200111107, 200111115\}@stu.hit.edu.cn} \\
  \texttt{\{pengchao.x, xujunjielong.l\}@bytedance.com} \quad \texttt{gaocuiyun@hit.edu.cn}
}

\begin{document}

\maketitle

\begin{abstract}

Scaling up executable code data
is significant for
improving language models' software engineering capability. The intricate nature of the process makes it labor-intensive, time-consuming, and expert-knowledge-dependent to build a large number of executable code repositories, limiting the scalability of existing work based on running tests. The primary bottleneck lies in the automated building of test environments for different repositories, which is an essential yet underexplored task.
To mitigate the gap, we introduce \tool, the first LLM-based agent aiming at automating
the building of executable test environments for any repositories at scale. Specifically, given a code repository, \tool~iteratively builds the Docker image, runs unit tests based on the feedback of the building, and synthesizes the Dockerfile until the entire pipeline is executed successfully.
The resulting Dockerfile can then be used to create Docker container environments for running code and tests.
We created a benchmark containing 420 Python repositories with unit tests for evaluation.
The results illustrate that \tool~achieves an 86.0\% success rate, outperforming SWE-agent by 77.0\%.
The resources of \tool~are available at \url{https://github.com/bytedance/Repo2Run}.

\end{abstract}

\section{Introduction}
\input{introduction}

\section{Formulation}
\input{configuration_definition}
\section{\tool}
\input{approach}

\section{Experiment}
\input{experiment}

\section{Discussion}
\input{discussion}

\section{Related Work}
\input{relatedworks}

\section{Conclusion}
\input{conclusion}

\section{Acknowledgement}
\input{acknowledgement}



\bibliography{Citation}

\newpage
\section*{NeurIPS Paper Checklist}

\input{checklist}
\newpage

\appendix
\input{appendix}

\end{document}

%% file: introduction.tex
\begin{figure}[htbp]
	\centering
	\includegraphics[width=\columnwidth]{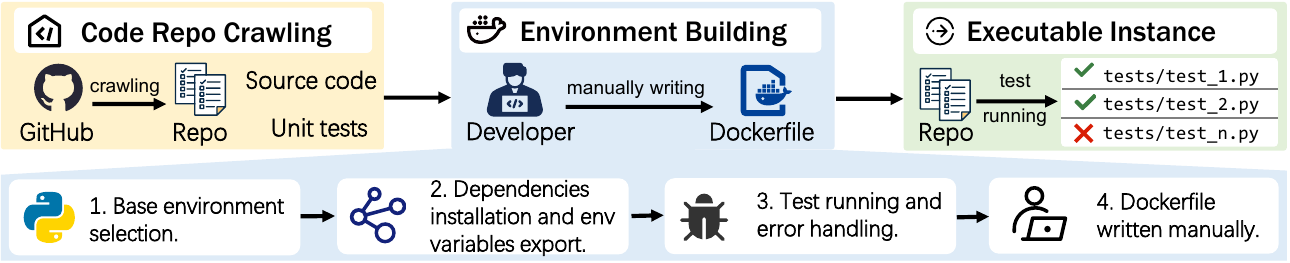}
    \caption{The pipeline of code repository mining and manual environment building. Developers manually write Dockerfiles through iterative steps including base environment selection, dependency installation,
    test running, error handling, validating the environment by running unit tests.} 
\label{figs:intro_new}
\end{figure}

\begin{figure}[htbp]
	\centering
	\includegraphics[width=\columnwidth]{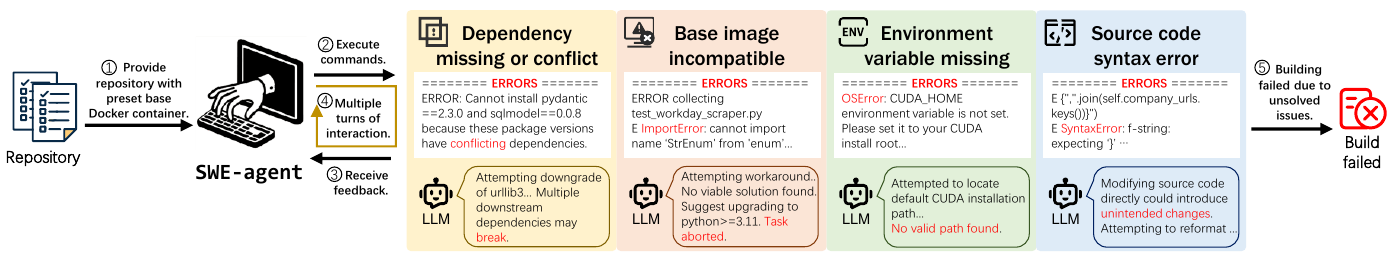}
    \caption{Four error types (highlighted in different colors)
    that SWE-agent fails
    to resolve during environment building.}
\label{figs:intro_challenge1}
\end{figure}

\begin{figure}[htbp]
	\centering
	\includegraphics[width=\columnwidth]{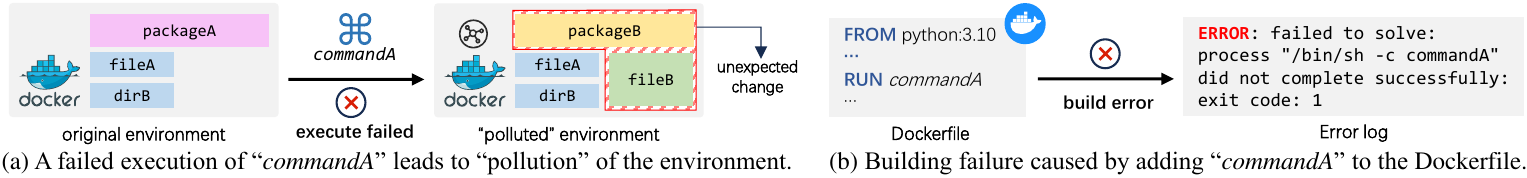}
    \caption{The illustration of a command executing failed and ``polluting'' the environment. (a) Failed commands like ``\textit{commandA}'' can irreversibly ``pollute'' the environment by altering packages, files, or directories, making subsequent builds unstable. (b) To reproduce the changes, such a failed command ``\texttt{RUN commandA}'' needs to be added to the Dockerfile. However, adding it will lead to building failure.}
\label{figs:intro_challenge2}
\end{figure}




Large language models (LLMs) have recently illustrated significant progress on solving software engineering issues~\cite{jimenez2024swe}, driving the advent of numerous coding LLM agents like MetaGPT~\cite{hong2024metagpt}, SWE-agent~\cite{yang2024swe}, OpenHands~\citep{wang2024openhands}, Copilot~\cite{copilot}, and Cursor~\cite{cursor}. 
However, the lack of training environment remains a significant challenge to advancing LLMs in software engineering — As shown in Figure~\ref{figs:intro_new}, while static code is readily accessible through the GitHub API, building \textit{executable environments} for code testing and patch validation demands extensive manual effort even for skilled developers~\footnote{We conducted a study to show how industry developers struggle in building environments. See Appendix~\ref{manual_experiment}.}.
This forces researchers to choose between (1) relying on unreliable static code metrics for rewarding (e.g., SWE-Fixer~\cite{xie2025swe}) and SWE-RL~\cite{wei2025swe}, and (2) engaging developers to manually build limited repository environments (e.g., SWE-Gym~\cite{pan2024training} and SWE-Smith~\cite{yang2025swe}) without scalability.
Thus, an automated environment building infrastructure is urgently needed but remains underexplored.


In this paper, we introduce \tool, an LLM-based agent designed to build executable environments.
The key idea is that if the agent can successfully navigate the process of building a repository's dependencies and running its unit tests, its actions can be recorded and replayed to programmatically synthesize a runnable Dockerfile.
This enables automated creation of isolated and consistent environments across different platforms via only one line command ``\texttt{docker build}'', eliminating ``\textit{it works on my machine}'' problem~\cite{valstar2020using}.

\tool~addresses two key challenges in automating executable environment building:

\begin{enumerate}[leftmargin=*]
    \item 
    \textbf{Hard to explore a valid trajectory that can successfully build the environment.} 
    We found that without effective actions, the LLM agent would struggle to handle such a complex process. As shown in Figure~\ref{figs:intro_challenge1}, SWE-agent~\cite{yang2024swe},
    a well-known LLM-based agent, often struggles with unresolved issues during environment building due to its lack of specialized actions.
    To address this challenge, \tool~uses an internal environment for building and an external environment for assistance, with nine actions designed to resolve these issues.
    \item 
    
    \textbf{Failing to synthesize a runnable Dockerfile without execution failure.}
    As shown in Figure~\ref{figs:intro_challenge2}, we observed that some incorrect commands in the trajectory can ``pollute'' the environment, leading to errors during synthesis.
    To tackle this challenge, we design an adaptive rollback action. When encountering failed commands like ``\textit{commandA}'' that may cause ``pollution'' to the environment, \tool~restores
    the internal environment to its pre-execution state. This ensures the building process remains consistent and repeatable using the Dockerfile, avoiding unexpected errors.
\end{enumerate}

To demonstrate \tool's effectiveness, we first created a benchmark of 420 latest Python repositories with unit tests from GitHub in 2024, as existing datasets include only a very small number of repositories~\cite{jimenez2024swe}. We choose Python for two key reasons: (1) the most popular software engineering benchmarks (e.g., SWE-bench~\cite{jimenez2024swe}) focus on Python repositories, and (2) Python is one of the most widely used programming languages today~\cite{githubreport}. We then evaluated \tool~on this benchmark, finding it successfully built the environments for 361 repositories, achieving an 86.0\% success rate, which is 63.9\% higher than the leading baseline. Notably, \tool~outperforms the SWE-agent by 77.0\%, demonstrating the effectiveness of the proposed agent.

We believe \tool~will serve as the foundational infrastructure, driving future research at the intersection of AI and Software Engineering and facilitating the community to efficiently scale up their executable code data and environments.

%% file: configuration_definition.tex

In this part, we define the task of executable environment building.
Given a repository $R$, executable environment building aims to determine a suitable \textbf{base image} $B$ and a \textbf{building process} $\mathcal{P}$ such that the resulting environment state $S$ satisfies the verification $\varepsilon$, as defined in Equation (\ref{eq1}):
\begin{equation}
    \varepsilon(R, S) = 0, \text{where}~S = \delta(B, \mathcal{P})
    \label{eq1}
\end{equation}
Here, $\delta$ is the state transition function, and $\varepsilon(R, S) = 0$ indicates that the verification $\varepsilon$ is satisfied.


\subsection{State transition}

$\bullet$ \textbf{Environment state ($S$):}
The environment state $S \in \mathcal{S}$ represents the current state of the computer system, which encompasses all variables, files, cache, etc.


$\bullet$ \textbf{Command sequence ($\mathcal{C}$):} The command sequence $\mathcal{C} \in \mathbb{C}$ represents a set of individual commands.
Each individual command $C \in \mathcal{C}$ refers to an instruction or action that can be executed in the environment via interfaces like \texttt{bash}, thereby changing the system state. 

$\bullet$ \textbf{State transition ($\delta$):}
The state transition $\delta$ is a function that defines the process through which the system transitions from the start state $S_{start} \in \mathcal{S}$ to the end state $S_{end} \in \mathcal{S}$ after executing of a command sequence $\mathcal{C}$, as defined in Equation (\ref{eq2}):
\begin{equation}
    \delta: \mathcal{S} \times \mathbb{C} \rightarrow \mathcal{S},\quad \delta(S_{start}, \mathcal{C}) = S_{end}
    \label{eq2}
\end{equation}

\subsection{Base image}

$\bullet$ \textbf{Empty state ($S_\emptyset$):}
The empty state $S_{\emptyset} \in \mathcal{S}$ represents a completely bare operating system or a purely hypothetical state without any builds.

$\bullet$ \textbf{Base image ($B$):}
The base image $B \in \mathcal{S}$ is a special type of environment state, typically managed by professional teams for user convenience, such as ``\texttt{python:3.10}''. Starting from the empty state $S_\emptyset$, the base image $B$ can be created by executing a predefined command sequence $\mathcal{C}_{B} \in \mathbb{C}$, i.e.,
$B = \delta(S_{\emptyset}, \mathcal{C}_{B})$.  Users can utilize these base images by simply adding their names to the Dockerfiles.


\subsection{Building process and state verification}

$\bullet$ \textbf{Building process ($\mathcal{P}$):}
The building process $\mathcal{P} \in \mathbb{C}$ is the command sequence designed to build the environment from the base image $B$. We denote the resulting state as $S_{f} \in \mathcal{S}$, where $S_{f} = \delta(B, \mathcal{P})$.



$\bullet$ \textbf{State verification ($\varepsilon$):}
The state verification $\varepsilon$ is a Boolean function used to determine whether the resulting state $S_f$ successfully runs all tests in the repository $R$.
$\varepsilon(R, S_{f}) = 0$
indicates that all tests in the repository $R$ can be successfully run. Otherwise, at least one test fails.

%% file: approach.tex

\begin{figure}[t]
	\centering
	\includegraphics[width=1.0\textwidth]{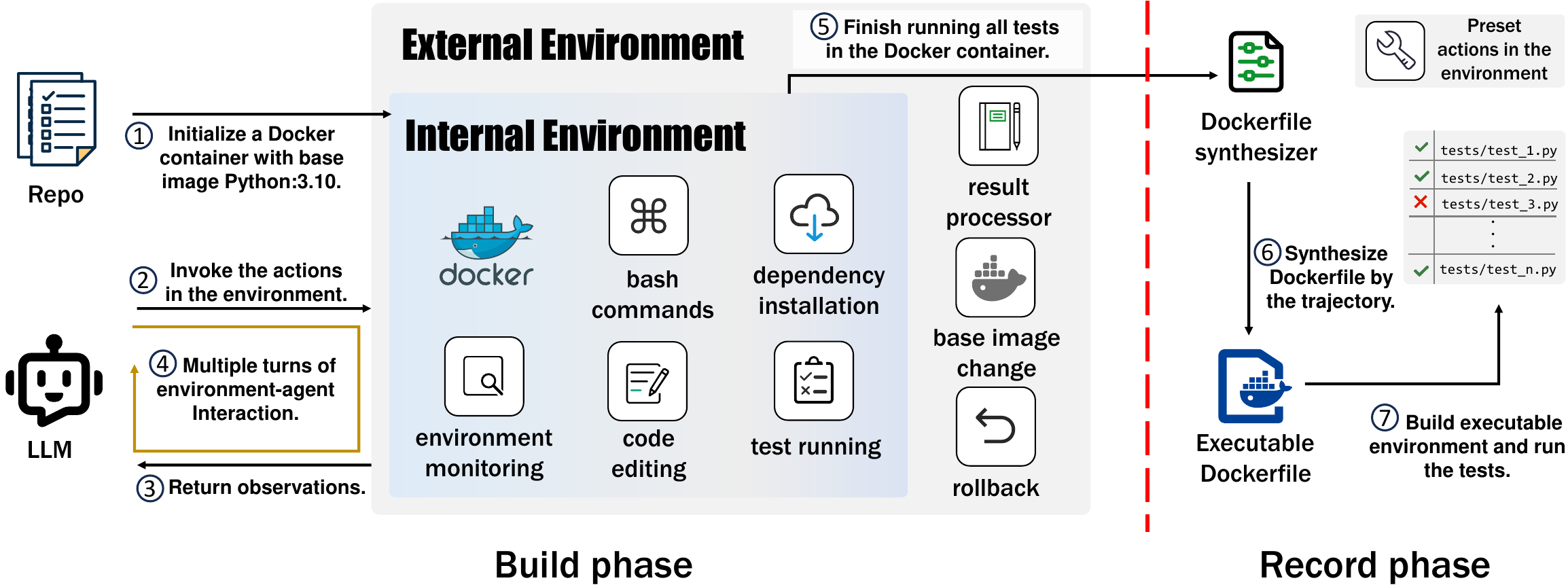}

    \caption{
    The workflow of \tool, involving two phases: the build phase and the record phase. The build phase utilizes a dual-environment architecture: the \textbf{internal environment} with five actions for environment building, while the \textbf{external environment} with three actions assists the internal environment.
    The record phase converts the validated command sequence into a runnable Dockerfile for reconstructing the executable environment. See Appendix~\ref{Repo2Run_example} for more examples of these actions.}
\label{figs:framework}
\end{figure}

In this part, we introduce the design of \tool.
Given a code repository, \tool~synthesizes a runnable Dockerfile, enabling the building of an executable environment. As shown in Figure~\ref{figs:framework}, this process is divided into two phases: the build phase and the record phase.


In \textbf{build phase}, \tool~employs a dual-environment architecture consisting of an \textbf{internal environment} and an \textbf{external environment}.
The internal environment is a Docker container where the agent can execute actions restricted to a container, e.g., test running.
The external environment is a terminal where the agent can execute actions prohibited within a container, e.g., base image change.
Once all tests in the internal environment are successfully executed, \tool~transitions to the record phase.

In \textbf{record phase}, \tool~synthesizes a runnable Dockerfile based on the command sequence executed during the build phase.
This Dockerfile serves as a precise record of the build process, and its execution can be viewed as a replay of the original build phase.

Next, we introduce the external environment, internal environment, and Dockerfile synthesizer.



\subsection{External environment}

The external environment serves as a bridge between the LLM and
internal environment. 
It transmits actions from the LLM to the internal environment and relaying observations back.
Following with the ReAct framework~\cite{DBLP:conf/iclr/YaoZYDSN023}, the external environment maintains a history of thoughts, actions (i.e., successfully executed commands), and observations. This iterative process ensures that the LLM stays informed about the system’s state and makes accurate decisions while effectively interacting with the internal environment.

$\bullet$ \textbf{Rollback}:
As shown in Figure~\ref{figs:intro_challenge2}, commands like ``\textit{commandA}''~\footnote{We collected some commands that can ``pollute'' the environment, see Appendix~\ref{pollution_commands}.} may ``pollute'' the environment, making it difficult to reproduce using a Dockerfile. When a command fails (i.e., returns a non-zero return code), the environment transitions to an uncertain state. Therefore, we introduce a rollback action to ensure the building process remains consistent.
Specifically, before executing a command, we use ``\texttt{docker commit}'' to create a reserve snapshot of the current state (i.e., save the image at that point). If the return code is not \texttt{0}, indicating a failure, the environment is rolled back to the most recent reserve snapshot. However, certain commands, referred to as safe commands (e.g., ``\texttt{cat}''), generally do not alter the state of the environment and are exempt from rollback. For a complete list of safe commands, see Appendix~\ref{safe_commands}.

$\bullet$ \textbf{Base image change}:
If the LLM agent determines that the current base image is unsuitable during the environment building process, it can reselect and switch to a new base image. This change 
invalidates the previous building process, clearing all executed commands and requiring a restart.

$\bullet$ \textbf{Result processor}: 
Command execution during interactions can produce extensive output (e.g., error logs), potentially overwhelming the LLM agent. To mitigate this, long outputs are truncated by this action, retaining only the initial and final sections up to a specified length.


\subsection{Internal environment}

The internal environment is a Docker container. Based on the latest data~\cite{python3data}
from 2025, we select \texttt{Python 3.10} as the default Docker base image due to its broadest adoption among Python versions. 

$\bullet$ \textbf{Environment monitoring}:
It serves as the eyes of the LLM agent in the internal environment, allowing it to observe the current environment state. 
These commands typically do not change the state of the environment. 
Basic commands like ``\texttt{ls}'' and ``\texttt{cat}'' are used to inspect directories and files, ``\texttt{find}'' is used to locate files, while more advanced commands like ``\texttt{pip list}'' and ``\texttt{pipdeptree}''~\cite{pipdeptree} retrieve the versions of installed libraries and dependency relationships.

$\bullet$ \textbf{Dependency installation}:
It serves as installing third-party packages required for running tests, using tools like \texttt{pip} for Python and \texttt{apt-get} for system-level packages. To resolve potential conflicts (e.g., version constraints), a dependency management is implemented to assist the installation, where packages to be installed are added to a waiting list and resolved as needed. If installation succeeds, the environment is updated; otherwise, the rollback is performed. For detailed designs, see Appendix~\ref{Repo2Run_tools}.


$\bullet$ \textbf{Test running}:
It serves as both a compass and a checkpoint, guiding the building process and verifying whether the environment runs all tests successfully. By executing ``\texttt{pytest}'' for unit tests, it determines whether the Docker container has been correctly built. If all tests run successfully, the process concludes; otherwise, error logs are sent to the LLM agent for further adjustments.

$\bullet$ \textbf{Code editing}:
It enables the LLM agent to modify the code within the internal environment, including both inside and outside the repository. Direct code editing is rare, but sometimes needed for issues like syntax errors. To prevent the LLM agent from bypassing tests by directly altering or deleting test files, it is restricted from modifying or deleting the original test files within the repository.

$\bullet$ \textbf{Bash commands}:
Like many other LLM-based agents~\cite{hong2024metagpt, yang2024swe, wang2024openhands}, \tool~is allowed to invoke bash commands, allowing it fully operate within the Docker container. It enables the agent to execute dynamically defined actions that are not pre-specified, ensuring flexibility in handling various errors.

\begin{figure}[t]
	\centering
	\includegraphics[width=1.0\textwidth]{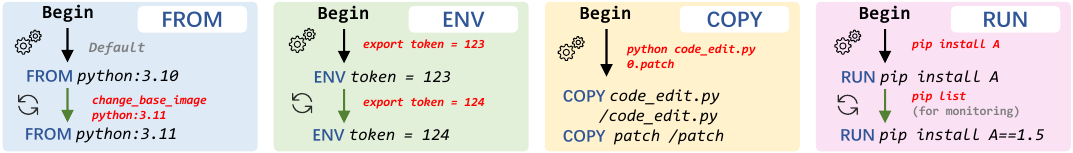}
    
    \caption{Rules for Dockerfile synthesis, illustrating how the Dockerfile synthesizer maps executed commands into Dockerfile statements using four keywords: ``\texttt{FROM}'', ``\texttt{ENV}'', ``\texttt{COPY}'', and ``\texttt{RUN}''. Black arrows represent the creation of statements, while green arrows indicate their transformations. Red text next to the arrows specifies the commands executed during each step.}
\label{figs:synthesis_rule}
\end{figure}

\subsection{Dockerfile synthesizer}
Once the LLM agent successfully runs all tests, the Dockerfile synthesizer converts the command sequence from the building process into a runnable Dockerfile. It processes each command sequentially, following the rules in Figure~\ref{figs:synthesis_rule} to synthesize Dockerfile statements. The synthesis uses four Dockerfile keywords: ``\texttt{FROM}'', ``\texttt{ENV}'', ``\texttt{COPY}'', and ``\texttt{RUN}''. For more details and examples, see Appendix~\ref{synthesizer_example}.

$\bullet$ \texttt{FROM}: The \texttt{FROM} statement defines the base image and is typically the first line in a Dockerfile. If the LLM agent changed the base image, the \texttt{FROM} statement must be updated, and all subsequents are cleared to ensure consistency.

$\bullet$ \texttt{ENV}: The \texttt{ENV} statement persistently sets environment variables in the Docker container. When a command with ``\texttt{export}'' is detected, it is converted into an \texttt{ENV} statement. If the vaule of an environment variable is overwritten, this statement is updated accordingly.

$\bullet$ \texttt{COPY}: The \texttt{COPY} statement copies local files or directories into the Docker container. As shown in Figure~\ref{figs:synthesis_rule}, it is often used for code editing to import editing scripts (i.e., \texttt{code\_edit.py}) and patches.

$\bullet$ \texttt{RUN}: The \texttt{RUN} statement executes commands in the container, with each line creating a new bash session. It cannot be used for persistent environment variables. Specially, for every installed packages, versions are recorded to ensure reproducibility, and the \texttt{RUN} statement is updated accordingly.


%% file: experiment.tex
We evaluate the effectiveness of \tool~on 420 Python code repositories.  As the popular option, we select \texttt{gpt-4o-2024-05-13} for all experiments, with the temperature uniformly set to 0.2.

\subsection{Benchmark}

To the best of our knowledge, there is no prior work similar to \tool~that automates executable environment building. Existing manually constructed datasets are limited to very few repositories~\cite{jimenez2024swe}.
To validate the capability of \tool, we create a new benchmark consisting of filtered Python repositories from GitHub based on the following criteria:

$\bullet$ \textbf{Creation date}: To avoid the data leakage, we select repositories created in 2024, ensuring they are not part of mainstream LLM training data.

$\bullet$ \textbf{Star count}: To maintain quality, we only include repositories with more than 100 stars.

$\bullet$ \textbf{Test directory}:
We focus on repositories likely to contain unit tests, identified using \texttt{pytest}, a leading Python testing framework compatible with tools like \texttt{unittest}. \texttt{Pytest} detects test files with a ``\texttt{test\_}'' prefix or ``\texttt{\_test}'' suffix, typically located in ``\texttt{test}'' or ``\texttt{tests}'' directories. We only retain repositories that have these directories.

Using these criteria, we crawled 449 repositories in December 2024 and filtered 420 containing at least one unit test to form our benchmark. For statistic of their scale, see Appendix~\ref{benchmark_statistic}.


\subsection{Evaluation metrics}

$\bullet$ \textbf{Dockerfile Generation Success Rate (DGSR)}:
It indicates the percentage of attempts where the method successfully generates a \textbf{runnable} Dockerfile. To be considered successful, the generated Dockerfile must be able to build without errors. If the Dockerfile for a code repository successfully builds, it is regarded as a successful generation. Generating runnable Dockerfile is fundamental for successfully building the executable environment.

$\bullet$ \textbf{Environment Building Success Rate (EBSR)}\footnote{We have also conducted a direct evaluation based on test pass rates in Appendix~\ref{pass_rate_experiment}.}:
It represents the percentage of attempts where the method successfully builds executable environments. For a successful building, the generated Dockerfile must not only build successfully but also allow tests to run by ``\texttt{pytest}'' in the Docker container. We are only concerned with whether tests can be executed, regardless of whether they pass or fail, as outcomes of tests may inherently vary within the repository.


\subsection{Baselines}

$\bullet$ \textbf{pipreqs}~\cite{pipreqs}: It is an automated tool that generates a ``\texttt{requirements.txt}'' file by analyzing the import statements in the Python scripts and identifying the necessary dependencies without LLM. Using the \texttt{requirements.txt} file generated by pipreqs, we create a Dockerfile. The detail is provided in Appendix~\ref{pipreqs_settings}.

$\bullet$ \textbf{LLM generator}:
The ``\texttt{README}'' file in a code repository usually contains environment building instructions. Therefore, we directly drive the LLM to read the ``\texttt{README}'' file and generate an executable Dockerfile accordingly.

$\bullet$ \textbf{SWE-agent}~\cite{yang2024swe}:
SWE-agent establishes a custom agent-computer interface (ACI) that uses the LLM agent's interaction with the repository environment by allowing actions such as reading files, editing files, and executing bash commands. Initially intended as an LLM agent for bug fixing, we preserve its framework and default settings, adjust its prompts, as shown in Appendix~\ref{swe_agent_setting}.

\subsection{Experimental Results}
\input{tables/compare}

\input{tables/different_type}

The results of different baselines are presented in Table \ref{tab:compare}. We observe that \tool~consistently outperforms other baselines on both DGSR and EBSR. Repo2Run ultimately completed environment building for 361 code repositories, achieving an EBSR of 86.0\%. It is 63.9\% higher than the highest rate achieved by other methods,  demonstrating great advantages.
Due to the design of rollback, \tool~successfully generates Dockerfiles that can be built successfully for all 420 code repositories, which other actions cannot guarantee. For details of each repostories, see Appendix~\ref{benchmark}.

For pipreqs, the main failures come from two reasons. First, generating the requirements.txt fails when there are issues within the repository, such as encoding errors or syntax errors in the files. This happens in 30 repositories (7.1\%). Second, even when requirements.txt is generated, it might not be downloaded
properly due to package version conflicts. This occurs in 265 repositories (63.1\%). Besides, both the LLM generator and SWE-agent fail to ensure that the generated Dockerfile can be successfully built due to the lack of an ensuring mechanism. Surprisingly, the ability of SWE-agent, a general agent framework, to generate Dockerfiles is even weaker than simply letting the LLM read the ``\texttt{README}'' file. This indicates that a general agent framework cannot guarantee the generation of runnable Dockerfiles. Ensuring mechanisms like rollback is necessary to effectively use the interactive information from the agent to generate runnable Dockerfiles.

As shown in Table~\ref{tab:different_types}, we manually categorized the 420 repositories into six domains. EBSR across these domains all exceed 80\%, outperforming all baselines and demonstrating \tool's consistent performance across different domains. Prior research~\cite{DBLP:conf/icse/HellendoornPGB19, DBLP:journals/corr/abs-2404-00599} shows that the real-world benchmarks differs from synthetic ones and better reflect a model's true capabilities. Our benchmark is built entirely from real-world GitHub data, filtered only by repository creation date, star count, and test, with no additional filtering. Thus, it accurately represents the current distribution of Python repositories. Moreover, a previous real-world benchmark~\cite{DBLP:journals/corr/abs-2404-00599} includes 59.6\% of AI/ML repositories, closely matching the 63.6\% proportion in our benchmark. This alignment further confirms that our benchmark reflects the real-world scenarios.

\subsection{Ablabtion of \tool}
\input{tables/ablation}
To investigate the impacts of the dual-environment architecture and Dockerfile synthesizer separately, as two parts of \tool, we separately remove each component of them. For the experiment without the dual-environment architecture, we retain only the internal environment's bash commands as the most basic interface and remove all other actions. For the experiment without the Dockerfile synthesizer, we directly instruct the LLM to synthesize a runnable Dockerfile. 

Experimental result of the ablation study is shown in Table~\ref{tab:ablation}. We observe that removing the dual-environment architecture and retaining only bash commands results in a 7.6\% decrease in DGSR. The main reason for this drop is the removal of rollback and other designs, making the system more prone to entering uncertain states and subsequently failing to reproduce. In addition, EBSR shows a 44.3\% decrease, primarily because the simplification of design makes it more difficult for the LLM agent to execute all tests in the internal environment. Besides, removing the Dockerfile synthesizer directly leads to an 80.5\% drop in DGSR. This indicates that having the LLM directly generate Dockerfiles is unlikely to fully follow the event history, resulting in Dockerfiles that fail to build successfully. This also directly causes a sharp decline in EBSR. Specially, we also perform an ablation study for the rollback. The results show that removing the rollback causes 3.1\% of generated Dockerfiles to become unrunnable, demonstrating the effectiveness of the rollback mechanism.

It is also observed that \tool~without the Dockerfile synthesizer is outperformed by the LLM generator. This is because the LLM generator leverages the ``\texttt{README}'' file, which provides a clear, simple and high-level overview of the executable environment building, allowing for more accurate Dockerfile synthesizer. In contrast, the event history-based approach lacks this context, making it harder for the LLM to fully understand the building goals. However, the Dockerfile synthesizer effectively utilizes the detailed event history, highlighting the complementary roles of both components of \tool~in generating a reliable runnable Dockerfile.



%% file: tables/compare.tex





\begin{table}[t]
    \small
    \setlength{\tabcolsep}{4.2mm} 
    \caption{The results of different baselines, including the percentage and number of successfully generated Dockerfiles and successfully built environments.}
    \centering
    \begin{tabular}{c c c c c}
    \toprule
    \rowcolor{gray!20}
    \textbf{Metric} & \textbf{DGSR} & \textbf{\makecell{\# Successfully \\ Generated Dockerfiles}} & \textbf{EBSR} & \textbf{\makecell{\# Successfully \\ Built Environments}} \\
    \midrule
    \textbf{pipreqs} & 29.8\% & 125 & 6.0\% & 25 \\
    \textbf{LLM generator} & 47.6\% & 200 & 22.1\% & 93 \\
    \textbf{SWE-agent} & 26.9\% & 113 & 9.0\% & 38 \\
    \midrule
    \textbf{\tool} & \textbf{100\%} & \textbf{420} & \textbf{86.0\%} & \textbf{361} \\
    \bottomrule
    \end{tabular}
    \label{tab:compare}
\end{table}

%% file: tables/different_type.tex
\begin{table}[htbp]
\centering
\setlength{\tabcolsep}{3.3mm}
\caption{Repository distribution across different domains and their EBSR, where SE means Software Engineering, NLP means Natural Language Processing, and DV means Data Visualization.}
\begin{tabular}{lccccccc}
\toprule
\rowcolor{gray!20} \textbf{Type} & \textbf{AI / ML} & \textbf{SE} & \textbf{NLP} & \textbf{DV} & \textbf{Security} & \textbf{Others} &\textbf{All} \\
\midrule
\textbf{\# Success} & 221 & 70 & 41 & 10 & 7 & 12 & 361 \\
\textbf{\# Total} & 267 & 73 & 47 & 11 & 8 & 14 & 420 \\
\midrule
\textbf{EBSR} & 82.8\% & 95.9\% & 87.2\% & 90.9\% & 87.5\% & 85.7\% & 86.0\% \\
\bottomrule
\end{tabular}
\label{tab:different_types}
\end{table}

%% file: tables/ablation.tex







\begin{table}[t]
    \small
    \setlength{\tabcolsep}{1mm}
    \caption{The results of ablation experiments, including the percentage and number of successfully generated Dockerfiles and successfully built environments.}
    \centering
    \begin{tabular}{cllcllc}
    \toprule
    \rowcolor{gray!20}
        \textbf{Metric} & \multicolumn{2}{c}{\textbf{DGSR}} & \textbf{\makecell{\# Successfully \\ Generated Dockerfiles}} & \multicolumn{2}{c}{\textbf{EBSR}} & \textbf{\makecell{\# Successfully \\ Built Environments}} \\
    \midrule
    \textbf{w/o dual-environment} &
    92.4\%  & \reddownarrow{7.6} & 388 & 
    41.7\%  & \reddownarrow{44.3} & 175\\
    $\hookrightarrow$ w/o rollback &
    96.9\%  & \reddownarrow{3.1} & 407 & 
    83.6\%  & \reddownarrow{2.4} & 351\\
    \textbf{w/o Dockerfile generator} &
    19.5\%  & \reddownarrow{80.5} & 82 & 
    13.8\%  & \reddownarrow{72.2} & 58\\
    \midrule
    \textbf{\tool} &
    \multicolumn{2}{c}{100\%} & 420 & 
    \multicolumn{2}{c}{86.0\%} & 361\\
    \bottomrule
    \end{tabular}
    \label{tab:ablation}
\end{table}

%% file: discussion.tex

\begin{wrapfigure}{r}{0.5\textwidth} 
    \centering
    \includegraphics[width=\linewidth]{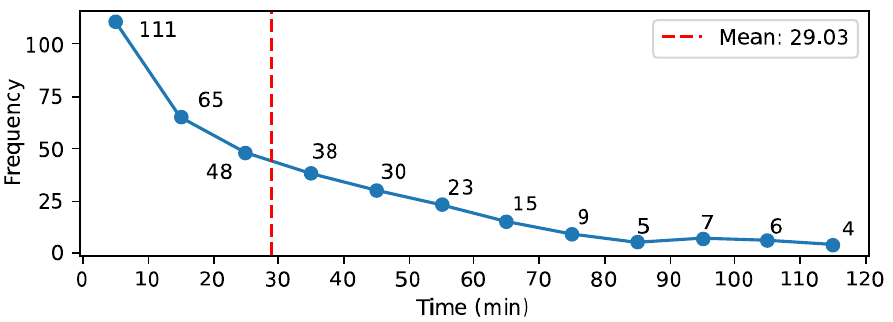} 
    \caption{Time distribution of successful building.}
    \label{figs:frequency}
\end{wrapfigure}


\subsection{Time consumption}


Figure~\ref{figs:frequency} shows the distribution of time spent successfully building 361 code repositories. The average time for successful building using \tool~is 29.03 minutes. 111 (30.7\%) of the repositories are successfully built in less than 10 minutes. Additionally, our empirical study through sampling indicates an average manual building time of 21.33 minutes (Appendix~\ref{manual_experiment}). Considering network differences and randomness, \tool~achieves a time consumption comparable to manual building. Additionally, for complex issues, \tool~shows greater advantages over manual building. \tool~successfully builds all the cases that were manually successful in our empirical study.

\noindent
\begin{minipage}[t]{0.48\textwidth}
\centering
\captionof{table}{Analysis of failed cases in \tool.}
\label{tab:failed_cases_analysis}
\setlength{\tabcolsep}{0.4mm}
\begin{tabular}{lr}
\toprule
\rowcolor{gray!20} \textbf{Category} & \textbf{\# Case (\%)} \\
\midrule
Hardware Insufficiency & 21 (35.6\%) \\
Missing Token & 1 (1.7\%) \\
Repository Defects & 5 (8.5\%) \\
Dependency Installation Timeout & 8 (13.5\%) \\
Runtest Timeout & 24 (40.7\%) \\
\bottomrule
\end{tabular}
\end{minipage}%
\hfill
\begin{minipage}[t]{0.48\textwidth}
\centering
\captionof{table}{Comparison of Dockerfile quality. ``Warning'' counts in each group show \tool~vs. baseline.}
\label{tab:dockerfile_quality_comparison}
\setlength{\tabcolsep}{4mm}
\begin{tabular}{lc}
\toprule
\rowcolor{gray!20} \textbf{Baseline} & \textbf{\# ``Warning''} \\
\midrule
vs. pipreqs & 5.17 / 6.0 \\
vs. LLM generator & 5.42 / 10.16 \\
vs. SWE-agent & 5.43 / 6.49 \\
\bottomrule
\end{tabular}
\end{minipage}

\subsection{Failure case study}


For the code repositories that fail to build, we manually inspect the reasons for failures and find that most are due to issues within the repositories themselves. 
Table~\ref{tab:failed_cases_analysis} demonstrates the analysis of 59 failed cases in \tool. The first three categories (45.8\%) primarily fail due to external factors, while the latter two categories (54.2\%) are unable to be configured successfully within the allotted time due to excessive dependencies and extensive tests inherent to the repositories. We also provide a detailed case in Appendix~\ref{failure_case}.

\subsection{Dockerfile Quality}



We utilize \texttt{hadolint}~\cite{hadolint} to analyze the quality of the Dockerfile generated by \tool, comparing with Dockerfiles successfully generated by three baselines and counting the number of warnings. Fewer warnings indicate higher quality in Dockerfile generation.
As shown in Table~\ref{tab:dockerfile_quality_comparison}, Dockerfiles generated by \tool’s consistently have fewer warnings than those generated by baselines.  Specifically, compared to \texttt{pipreqs}, LLM generator, and SWE-agent, \tool~reduces warnings from 6.0 to 5.17, 10.16 to 5.42, and 6.49 to 5.43, respectively. The results indicate that \tool~ produces higher-quality Dockerfiles with fewer potential issues.

\subsection{Limitation}
\label{limitation}
\tool~can automate building executable environments for many repositories but may still encounter challenges in some edge cases. For complex environments that remain unresolved, manual verification and intervention are necessary. Nonetheless, as an important step toward full automation, we believe this work makes a significant contribution to the AI community’s infrastructure.

%% file: relatedworks.tex
\subsection{LLM-based agent}


LLM-based agents~\cite{liu2024marscode, gao2025trae} typically consist of four key components: planning, memory, perception, and action~\cite{components}. Planning is crucial for agent systems, as it schedules agents to ensure a smooth process. LLM-based agents employ various planning strategies, including single~\cite{plan1} or multiple planners~\cite{plan2}, single~\cite{plan3} or multi-turn planning~\cite{plan4}, and single~\cite{plan5} or multi-path planning~\cite{plan6}. The memory component in LLM-based agents stores historical data to support coherent reasoning and complex tasks. Implementations vary in terms of memory duration (short-term~\cite{memory1} or long-term memory~\cite{memory2}), memory ownership (specific~\cite{memory3} or shared memory~\cite{memory4}).
For perception, LLM-based agents primarily utilize textual input~\cite{perception1, memory2} (natural language and programming language) and visual input~\cite{perception3, perception4} (images and diagrams) to perceive and process information.
To extend capabilities beyond interactive dialogue, the action component employs various external tools~\cite{action1, action2}, such as searching tools, file operations, and GUI operations. 
Nowadays, LLM-based agents have demonstrated superior performance compared to standalone LLMs in various software engineering tasks~\cite{agent1,aegis,agent2, agent3,wang2025repomaster,wang2025codevisionary,xu2025openrca,wen2025vul,ni2025gittaskbench,xu2024aligning}. Many complex software engineering problems, especially repository-level challenges~\cite{DBLP:conf/acl/WenYGXY25,wang2024reposvul,wu2024repomastereval,DBLP:journals/corr/abs-2404-15596}, can thereby be effectively addressed.
However, no LLM-based agents are specifically designed for executable environment building currently. To fill this gap, this paper employs a novel approach for automated executable coding environment building and runnable Dockerfile generation. 


\subsection{Environment building and Dockerfile generation}

There have already been many efforts devoted to environment building. Oss-Fuzz-Gen~\cite{Liu_OSS-Fuzz-Gen_Automated_Fuzz_2024} relies on predefined build instructions (such as ``\texttt{./bootstrap.sh}'', ``\texttt{./configure}'', ``\texttt{make}'') to build projects for fuzzing but lacks of flexibility for diverse projects when specified files are absent.
Existing solutions that help developers write Dockerfiles broadly fall into three categories: (1) Template-based generators that create Dockerfiles based on project context~\cite{starter, generatordocker}, (2) Task-specific tools such as pipreqs~\cite{pipreqs}, DockerizeMe~\cite{horton2019dockerizeme} which supports environment dependency inference for Python projects, and DockerGen~\cite{ye2021dockergen} for dependency recommendations based on knowledge graphs built from existing Dockerfiles, whose overall effectiveness still has substantial room for improvement~\cite{peng2024less}, and (3) Code completion tools including GitHub Copilot~\cite{copilot} and HumpBack~\cite{hanayama2020humpback} that generate suggestions for developers while writing Dockerfiles. The use of deep learning models to generate Dockerfiles based on natural language specifications of software requirements is also investigated~\cite{rosa2023automatically}. While these approaches provide valuable assistance, they either require significant manual input from developers or are limited to specific use cases.

%% file: conclusion.tex
In this paper, we propose \tool, the first LLM-based agent for automated coding executable environment building and Dockerfile generation for Python repositories. With a dual-environment architecture and a Dockerfile synthesizer, \tool~is able to select and change the base image, manage and install dependencies based on action observation and rollback mechanism, and utilize bash commands and existing test suites. Our evaluation of 420 popular Python repositories hosted on GitHub demonstrates the effectiveness of \tool~with an 86.0\% success rate. We believe \tool~will serve as the foundational infrastructure, enabling the community to efficiently scale up executable environments.

%% file: acknowledgement.tex
This research is supported by the National Natural Science Foundation of China under project (No. 62472126), Natural Science Foundation of Guangdong Province (Project No. 2023A1515011959), and Shenzhen-Hong Kong Jointly Funded Project (Category A, No. SGDX20230116 091246007).

%% file: checklist.tex
\begin{enumerate}

\item {\bf Claims}
    \item[] Question: Do the main claims made in the abstract and introduction accurately reflect the paper's contributions and scope?
    \item[] Answer: \answerYes{} 
    \item[] Justification: The abstract and introduction clearly state the main claims, contributions, and scope of the paper.
    \item[] Guidelines:
    \begin{itemize}
        \item The answer NA means that the abstract and introduction do not include the claims made in the paper.
        \item The abstract and/or introduction should clearly state the claims made, including the contributions made in the paper and important assumptions and limitations. A No or NA answer to this question will not be perceived well by the reviewers. 
        \item The claims made should match theoretical and experimental results, and reflect how much the results can be expected to generalize to other settings. 
        \item It is fine to include aspirational goals as motivation as long as it is clear that these goals are not attained by the paper. 
    \end{itemize}

\item {\bf Limitations}
    \item[] Question: Does the paper discuss the limitations of the work performed by the authors?
    \item[] Answer: \answerYes{} 
    \item[] Justification: The paper includes a dedicated ``Limitations'' section in Section~\ref{limitation}. We also reflect on the impact of these limitations on the results and their generalizability.
    \item[] Guidelines:
    \begin{itemize}
        \item The answer NA means that the paper has no limitation while the answer No means that the paper has limitations, but those are not discussed in the paper. 
        \item The authors are encouraged to create a separate "Limitations" section in their paper.
        \item The paper should point out any strong assumptions and how robust the results are to violations of these assumptions (e.g., independence assumptions, noiseless settings, model well-specification, asymptotic approximations only holding locally). The authors should reflect on how these assumptions might be violated in practice and what the implications would be.
        \item The authors should reflect on the scope of the claims made, e.g., if the approach was only tested on a few datasets or with a few runs. In general, empirical results often depend on implicit assumptions, which should be articulated.
        \item The authors should reflect on the factors that influence the performance of the approach. For example, a facial recognition algorithm may perform poorly when image resolution is low or images are taken in low lighting. Or a speech-to-text system might not be used reliably to provide closed captions for online lectures because it fails to handle technical jargon.
        \item The authors should discuss the computational efficiency of the proposed algorithms and how they scale with dataset size.
        \item If applicable, the authors should discuss possible limitations of their approach to address problems of privacy and fairness.
        \item While the authors might fear that complete honesty about limitations might be used by reviewers as grounds for rejection, a worse outcome might be that reviewers discover limitations that aren't acknowledged in the paper. The authors should use their best judgment and recognize that individual actions in favor of transparency play an important role in developing norms that preserve the integrity of the community. Reviewers will be specifically instructed to not penalize honesty concerning limitations.
    \end{itemize}

\item {\bf Theory assumptions and proofs}
    \item[] Question: For each theoretical result, does the paper provide the full set of assumptions and a complete (and correct) proof?
    \item[] Answer: \answerNA{} 
    \item[] Justification: The paper does not include any theoretical results, so this question is not applicable.
    \item[] Guidelines:
    \begin{itemize}
        \item The answer NA means that the paper does not include theoretical results. 
        \item All the theorems, formulas, and proofs in the paper should be numbered and cross-referenced.
        \item All assumptions should be clearly stated or referenced in the statement of any theorems.
        \item The proofs can either appear in the main paper or the supplemental material, but if they appear in the supplemental material, the authors are encouraged to provide a short proof sketch to provide intuition. 
        \item Inversely, any informal proof provided in the core of the paper should be complemented by formal proofs provided in appendix or supplemental material.
        \item Theorems and Lemmas that the proof relies upon should be properly referenced. 
    \end{itemize}

    \item {\bf Experimental result reproducibility}
    \item[] Question: Does the paper fully disclose all the information needed to reproduce the main experimental results of the paper to the extent that it affects the main claims and/or conclusions of the paper (regardless of whether the code and data are provided or not)?
    \item[] Answer: \answerYes{} 
    \item[] Justification: We provide the complete code along with instructions to build an executable environment, ensuring that the main experimental results can be fully reproduced. This allows others to verify and replicate our findings reliably.
    \item[] Guidelines:
    \begin{itemize}
        \item The answer NA means that the paper does not include experiments.
        \item If the paper includes experiments, a No answer to this question will not be perceived well by the reviewers: Making the paper reproducible is important, regardless of whether the code and data are provided or not.
        \item If the contribution is a dataset and/or model, the authors should describe the steps taken to make their results reproducible or verifiable. 
        \item Depending on the contribution, reproducibility can be accomplished in various ways. For example, if the contribution is a novel architecture, describing the architecture fully might suffice, or if the contribution is a specific model and empirical evaluation, it may be necessary to either make it possible for others to replicate the model with the same dataset, or provide access to the model. In general. releasing code and data is often one good way to accomplish this, but reproducibility can also be provided via detailed instructions for how to replicate the results, access to a hosted model (e.g., in the case of a large language model), releasing of a model checkpoint, or other means that are appropriate to the research performed.
        \item While NeurIPS does not require releasing code, the conference does require all submissions to provide some reasonable avenue for reproducibility, which may depend on the nature of the contribution. For example
        \begin{enumerate}
            \item If the contribution is primarily a new algorithm, the paper should make it clear how to reproduce that algorithm.
            \item If the contribution is primarily a new model architecture, the paper should describe the architecture clearly and fully.
            \item If the contribution is a new model (e.g., a large language model), then there should either be a way to access this model for reproducing the results or a way to reproduce the model (e.g., with an open-source dataset or instructions for how to construct the dataset).
            \item We recognize that reproducibility may be tricky in some cases, in which case authors are welcome to describe the particular way they provide for reproducibility. In the case of closed-source models, it may be that access to the model is limited in some way (e.g., to registered users), but it should be possible for other researchers to have some path to reproducing or verifying the results.
        \end{enumerate}
    \end{itemize}

\item {\bf Open access to data and code}
    \item[] Question: Does the paper provide open access to the data and code, with sufficient instructions to faithfully reproduce the main experimental results, as described in supplemental material?
    \item[] Answer: \answerYes{} 
    \item[] Justification: We provide the code and data in our anonymized repository.
    \item[] Guidelines:
    \begin{itemize}
        \item The answer NA means that paper does not include experiments requiring code.
        \item Please see the NeurIPS code and data submission guidelines (\url{https://nips.cc/public/guides/CodeSubmissionPolicy}) for more details.
        \item While we encourage the release of code and data, we understand that this might not be possible, so “No” is an acceptable answer. Papers cannot be rejected simply for not including code, unless this is central to the contribution (e.g., for a new open-source benchmark).
        \item The instructions should contain the exact command and environment needed to run to reproduce the results. See the NeurIPS code and data submission guidelines (\url{https://nips.cc/public/guides/CodeSubmissionPolicy}) for more details.
        \item The authors should provide instructions on data access and preparation, including how to access the raw data, preprocessed data, intermediate data, and generated data, etc.
        \item The authors should provide scripts to reproduce all experimental results for the new proposed method and baselines. If only a subset of experiments are reproducible, they should state which ones are omitted from the script and why.
        \item At submission time, to preserve anonymity, the authors should release anonymized versions (if applicable).
        \item Providing as much information as possible in supplemental material (appended to the paper) is recommended, but including URLs to data and code is permitted.
    \end{itemize}

\item {\bf Experimental setting/details}
    \item[] Question: Does the paper specify all the training and test details (e.g., data splits, hyperparameters, how they were chosen, type of optimizer, etc.) necessary to understand the results?
    \item[] Answer: \answerYes{} 
    \item[] Justification: We describe the experimental design in the main paper and provide the LLM prompts in an anonymized repository to ensure full transparency and reproducibility.
    \item[] Guidelines:
    \begin{itemize}
        \item The answer NA means that the paper does not include experiments.
        \item The experimental setting should be presented in the core of the paper to a level of detail that is necessary to appreciate the results and make sense of them.
        \item The full details can be provided either with the code, in appendix, or as supplemental material.
    \end{itemize}

\item {\bf Experiment statistical significance}
    \item[] Question: Does the paper report error bars suitably and correctly defined or other appropriate information about the statistical significance of the experiments?
    \item[] Answer: \answerNo{} 
    \item[] Justification: We have not repeated experiments many times to get error bars since experiments of multiple times is time-consuming and expensive.
    \item[] Guidelines:
    \begin{itemize}
        \item The answer NA means that the paper does not include experiments.
        \item The authors should answer "Yes" if the results are accompanied by error bars, confidence intervals, or statistical significance tests, at least for the experiments that support the main claims of the paper.
        \item The factors of variability that the error bars are capturing should be clearly stated (for example, train/test split, initialization, random drawing of some parameter, or overall run with given experimental conditions).
        \item The method for calculating the error bars should be explained (closed form formula, call to a library function, bootstrap, etc.)
        \item The assumptions made should be given (e.g., Normally distributed errors).
        \item It should be clear whether the error bar is the standard deviation or the standard error of the mean.
        \item It is OK to report 1-sigma error bars, but one should state it. The authors should preferably report a 2-sigma error bar than state that they have a 96\% CI, if the hypothesis of Normality of errors is not verified.
        \item For asymmetric distributions, the authors should be careful not to show in tables or figures symmetric error bars that would yield results that are out of range (e.g. negative error rates).
        \item If error bars are reported in tables or plots, The authors should explain in the text how they were calculated and reference the corresponding figures or tables in the text.
    \end{itemize}

\item {\bf Experiments compute resources}
    \item[] Question: For each experiment, does the paper provide sufficient information on the computer resources (type of compute workers, memory, time of execution) needed to reproduce the experiments?
    \item[] Answer: \answerYes{} 
    \item[] Justification: We specify the LLM used in the experiments and report the temperature settings applied during training and evaluation.
    \item[] Guidelines:
    \begin{itemize}
        \item The answer NA means that the paper does not include experiments.
        \item The paper should indicate the type of compute workers CPU or GPU, internal cluster, or cloud provider, including relevant memory and storage.
        \item The paper should provide the amount of compute required for each of the individual experimental runs as well as estimate the total compute. 
        \item The paper should disclose whether the full research project required more compute than the experiments reported in the paper (e.g., preliminary or failed experiments that didn't make it into the paper). 
    \end{itemize}
    
\item {\bf Code of ethics}
    \item[] Question: Does the research conducted in the paper conform, in every respect, with the NeurIPS Code of Ethics \url{https://neurips.cc/public/EthicsGuidelines}?
    \item[] Answer: \answerYes{} 
    \item[] Justification: Our research fully complies with the NeurIPS Code of Ethics in all respects.
    \item[] Guidelines:
    \begin{itemize}
        \item The answer NA means that the authors have not reviewed the NeurIPS Code of Ethics.
        \item If the authors answer No, they should explain the special circumstances that require a deviation from the Code of Ethics.
        \item The authors should make sure to preserve anonymity (e.g., if there is a special consideration due to laws or regulations in their jurisdiction).
    \end{itemize}

\item {\bf Broader impacts}
    \item[] Question: Does the paper discuss both potential positive societal impacts and negative societal impacts of the work performed?
    \item[] Answer: \answerNA{} 
    \item[] Justification: The research presented in this paper remains at the experimental levels and has not yet resulted in any societal impacts.
    \item[] Guidelines:
    \begin{itemize}
        \item The answer NA means that there is no societal impact of the work performed.
        \item If the authors answer NA or No, they should explain why their work has no societal impact or why the paper does not address societal impact.
        \item Examples of negative societal impacts include potential malicious or unintended uses (e.g., disinformation, generating fake profiles, surveillance), fairness considerations (e.g., deployment of technologies that could make decisions that unfairly impact specific groups), privacy considerations, and security considerations.
        \item The conference expects that many papers will be foundational research and not tied to particular applications, let alone deployments. However, if there is a direct path to any negative applications, the authors should point it out. For example, it is legitimate to point out that an improvement in the quality of generative models could be used to generate deepfakes for disinformation. On the other hand, it is not needed to point out that a generic algorithm for optimizing neural networks could enable people to train models that generate Deepfakes faster.
        \item The authors should consider possible harms that could arise when the technology is being used as intended and functioning correctly, harms that could arise when the technology is being used as intended but gives incorrect results, and harms following from (intentional or unintentional) misuse of the technology.
        \item If there are negative societal impacts, the authors could also discuss possible mitigation strategies (e.g., gated release of models, providing defenses in addition to attacks, mechanisms for monitoring misuse, mechanisms to monitor how a system learns from feedback over time, improving the efficiency and accessibility of ML).
    \end{itemize}
    
\item {\bf Safeguards}
    \item[] Question: Does the paper describe safeguards that have been put in place for responsible release of data or models that have a high risk for misuse (e.g., pretrained language models, image generators, or scraped datasets)?
    \item[] Answer: \answerNA{} 
    \item[] Justification: Our work does not pose any risks related to misuse or dual-use, so no special safeguards are necessary.
    \item[] Guidelines:
    \begin{itemize}
        \item The answer NA means that the paper poses no such risks.
        \item Released models that have a high risk for misuse or dual-use should be released with necessary safeguards to allow for controlled use of the model, for example by requiring that users adhere to usage guidelines or restrictions to access the model or implementing safety filters. 
        \item Datasets that have been scraped from the Internet could pose safety risks. The authors should describe how they avoided releasing unsafe images.
        \item We recognize that providing effective safeguards is challenging, and many papers do not require this, but we encourage authors to take this into account and make a best faith effort.
    \end{itemize}

\item {\bf Licenses for existing assets}
    \item[] Question: Are the creators or original owners of assets (e.g., code, data, models), used in the paper, properly credited and are the license and terms of use explicitly mentioned and properly respected?
    \item[] Answer: \answerYes{} 
    \item[] Justification: We have properly cited all original creators and clearly stated the licenses and terms of use for all assets used in our paper. In the implementation of \tool, dependencies are clearly cited in the ``\texttt{requirements.txt}''. The usage of open-source code complies to its license and the copyright notice and source are placed in the our source code.
    \item[] Guidelines:
    \begin{itemize}
        \item The answer NA means that the paper does not use existing assets.
        \item The authors should cite the original paper that produced the code package or dataset.
        \item The authors should state which version of the asset is used and, if possible, include a URL.
        \item The name of the license (e.g., CC-BY 4.0) should be included for each asset.
        \item For scraped data from a particular source (e.g., website), the copyright and terms of service of that source should be provided.
        \item If assets are released, the license, copyright information, and terms of use in the package should be provided. For popular datasets, \url{paperswithcode.com/datasets} has curated licenses for some datasets. Their licensing guide can help determine the license of a dataset.
        \item For existing datasets that are re-packaged, both the original license and the license of the derived asset (if it has changed) should be provided.
        \item If this information is not available online, the authors are encouraged to reach out to the asset's creators.
    \end{itemize}

\item {\bf New assets}
    \item[] Question: Are new assets introduced in the paper well documented and is the documentation provided alongside the assets?
    \item[] Answer: \answerYes{} 
    \item[] Justification: Our benchmark is constructed in accordance with the relevant license requirements, ensuring proper compliance and documentation. The usage of the implementation of \tool~is provided in its \texttt{README.md} file.
    \item[] Guidelines:
    \begin{itemize}
        \item The answer NA means that the paper does not release new assets.
        \item Researchers should communicate the details of the dataset/code/model as part of their submissions via structured templates. This includes details about training, license, limitations, etc. 
        \item The paper should discuss whether and how consent was obtained from people whose asset is used.
        \item At submission time, remember to anonymize your assets (if applicable). You can either create an anonymized URL or include an anonymized zip file.
    \end{itemize}

\item {\bf Crowdsourcing and research with human subjects}
    \item[] Question: For crowdsourcing experiments and research with human subjects, does the paper include the full text of instructions given to participants and screenshots, if applicable, as well as details about compensation (if any)? 
    \item[] Answer: \answerYes{} 
    \item[] Justification: We provided the full instructions and details for the study on how industry developers struggle in building environments in Section~\ref{manual_experiment}. The human study was approved by the team leader and legal business partner of these developers, and was conducted during their working hours as paid work. Their salary was above the minimum wage in the country where this study was conducted.
    No compensation details are involved as participants were not compensated.
    \item[] Guidelines:
    \begin{itemize}
        \item The answer NA means that the paper does not involve crowdsourcing nor research with human subjects.
        \item Including this information in the supplemental material is fine, but if the main contribution of the paper involves human subjects, then as much detail as possible should be included in the main paper. 
        \item According to the NeurIPS Code of Ethics, workers involved in data collection, curation, or other labor should be paid at least the minimum wage in the country of the data collector. 
    \end{itemize}

\item {\bf Institutional review board (IRB) approvals or equivalent for research with human subjects}
    \item[] Question: Does the paper describe potential risks incurred by study participants, whether such risks were disclosed to the subjects, and whether Institutional Review Board (IRB) approvals (or an equivalent approval/review based on the requirements of your country or institution) were obtained?
    \item[] Answer: \answerYes{} 
    \item[] Justification: Our study on how industry developers struggle in building environments in Section~\ref{manual_experiment} involved only survey participation and was approved by the IRB from their organization. We have complied with all relevant ethical guidelines.
    \item[] Guidelines:
    \begin{itemize}
        \item The answer NA means that the paper does not involve crowdsourcing nor research with human subjects.
        \item Depending on the country in which research is conducted, IRB approval (or equivalent) may be required for any human subjects research. If you obtained IRB approval, you should clearly state this in the paper. 
        \item We recognize that the procedures for this may vary significantly between institutions and locations, and we expect authors to adhere to the NeurIPS Code of Ethics and the guidelines for their institution. 
        \item For initial submissions, do not include any information that would break anonymity (if applicable), such as the institution conducting the review.
    \end{itemize}

\item {\bf Declaration of LLM usage}
    \item[] Question: Does the paper describe the usage of LLMs if it is an important, original, or non-standard component of the core methods in this research? Note that if the LLM is used only for writing, editing, or formatting purposes and does not impact the core methodology, scientific rigorousness, or originality of the research, declaration is not required.
    \item[] Answer: \answerYes{} 
    \item[] Justification: We provide a thorough description of the usage of LLMs as a core component of our research.
    \item[] Guidelines:
    \begin{itemize}
        \item The answer NA means that the core method development in this research does not involve LLMs as any important, original, or non-standard components.
        \item Please refer to our LLM policy (\url{https://neurips.cc/Conferences/2025/LLM}) for what should or should not be described.
    \end{itemize}

\end{enumerate}

%% file: appendix.tex
\newpage
\begin{center}
    \LARGE \textbf{Appendix}
\end{center}

\section{\tool~details}
\label{Repo2Run_example}
\input{appendix/Repo2Run_example}

\section{Dockerfile synthesizer details}
\label{synthesizer_example}
\input{appendix/synthesizer_example}

\section{Pollution commands}
\label{pollution_commands}
\input{appendix/pollution_commands}
\section{Benchmark statistic}
\label{benchmark_statistic}
\input{appendix/benchmark_statistic}

\section{Test pass rate experiment}
\label{pass_rate_experiment}
\input{appendix/pass_rate_experiment}

\section{Empirical study}
\label{manual_experiment}
\input{appendix/manual_experiment}


\section{\tool~tools}
\label{Repo2Run_tools}
\input{appendix/Repo2Run_tools}

\newpage
\section{Repository building success status}
\label{benchmark}
\input{tables/benchmark}

\newpage
\section{Baseline settings}
\label{baseline_settings}
\input{appendix/baseline_settings}

\section{Failure Case}
\label{failure_case}
\input{appendix/failure_case}

%% file: appendix/Repo2Run_example.tex
\subsection{Example of \tool}
\begin{figure}[hbtp]
	\centering
	\includegraphics[width=1.0\textwidth]{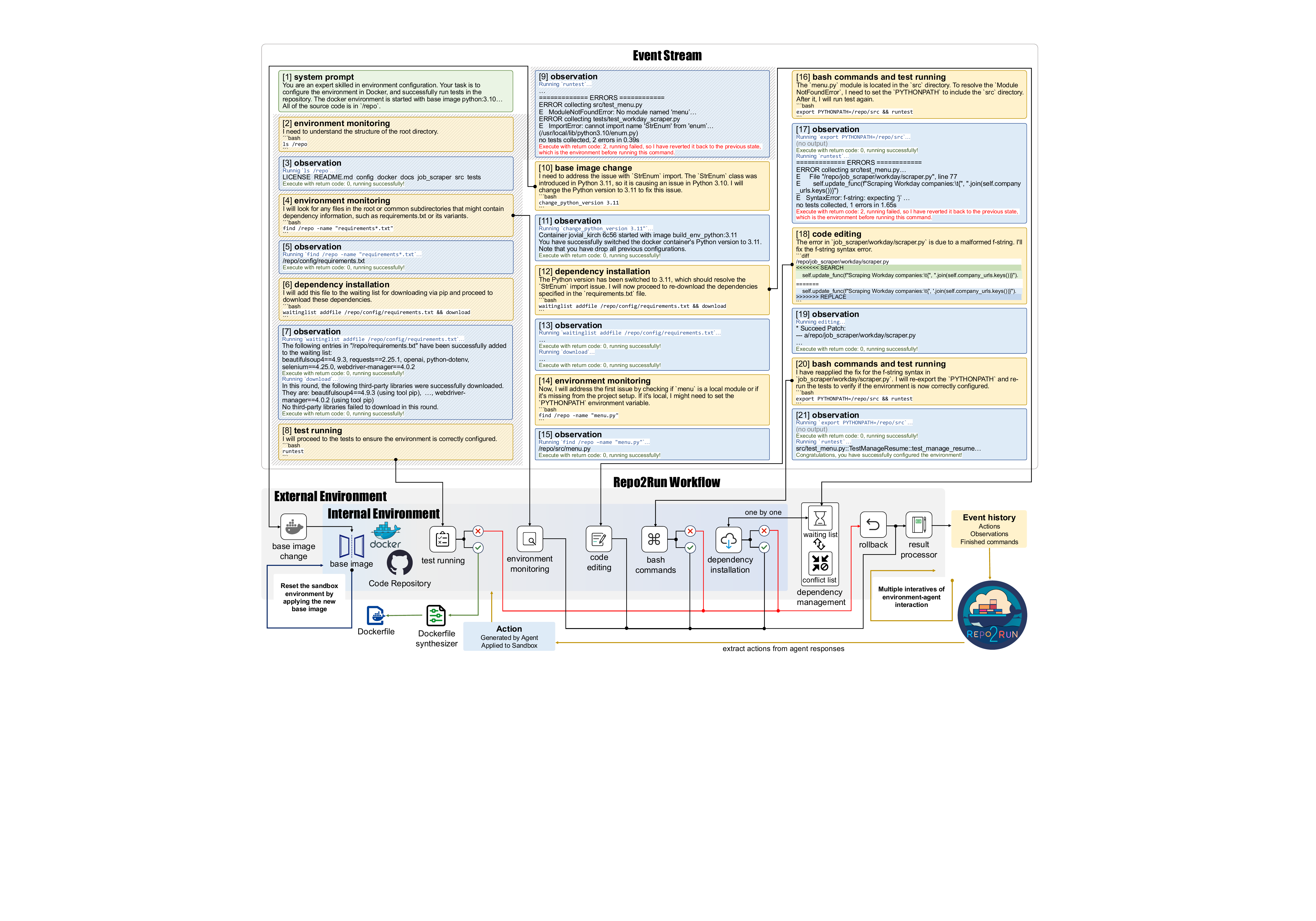}

    \caption{An example process of \textbf{\tool}, which illustrates two main parts: \textbf{Event Stream} and \textbf{\tool}~\textbf{Workflow}. The Event Stream tracks the entire action-observation process, where the green box represents the abbreviated system prompt, yellow boxes represent responses from the LLM agent including the actions, and blue boxes represent observations from the Runtime Environment. The shaded actions indicate buildings abandoned after \textbf{base image change}. In the Event Stream, the blue text indicates that the command starts running, the green text indicates that the command runs successfully, and the red text indicates that the command fails. The \textbf{\tool}~\textbf{Workflow} consists of the \textbf{internal environment} and the \textbf{external environment}. The internal environment serves as the actual buildings Docker-based sandbox, which builds an actual testing runtime environment. The external environment executes the action-observation process and assists in the building process within the internal environment.}
\label{figs:framework_old}
\end{figure}

Figure~\ref{figs:framework_old} shows an example process of \tool, including the event stream and \tool~ workflow. Below, we introduce the executed action based on the example:

\begin{itemize}[left=0pt, label=\textbullet, itemsep=0em] 
  \item \textbf{Base Image Change}: In the tenth part of Event Stream, when the LLM
 agent discovers that the ``\texttt{}{StrEnum}'' class cannot be imported through testing, which is only available in the Python
 standard library starting from version 3.11, it decides to
 reselect the base image to python:3.11.
  \item \textbf{Environment monitoring}: In the fourth part, the LLM agent calls the ``\texttt{find}'' command to search for \texttt{requirements.txt} within the environment to determine which dependencies to install.
  \item \textbf{Dependency installation}: In the sixth part, the action ``\texttt{waitinglist addfile}'' adds all elements from \texttt{requirements.txt} to the waiting list, while executing the ``\texttt{download}'' command. Third-party libraries are taken one by one from the waiting list and installed in the internal environment. The installation commands (e.g., ``\texttt{pip install}'' and ``\texttt{apt-get install}'') are executed. 
  \item \textbf{Code editing}: The eighteenth part demonstrates code editing to correct a syntax error resulting from the improper use of double quotes in a Python f-string.
  \item \textbf{Bash commands}: The sixteenth part shows that the LLM agent uses the command ``\texttt{export PYTHONPATH=/repo/src}''.
  \item \textbf{Test running}: The eighth part shows that the LLM agent calls the ``\texttt{runtest}'' command to execute unit tests.
\end{itemize}

\subsection{Action usage frequency}
\begin{figure}[hbtp]
	\centering
	\includegraphics[width=1.0\textwidth]{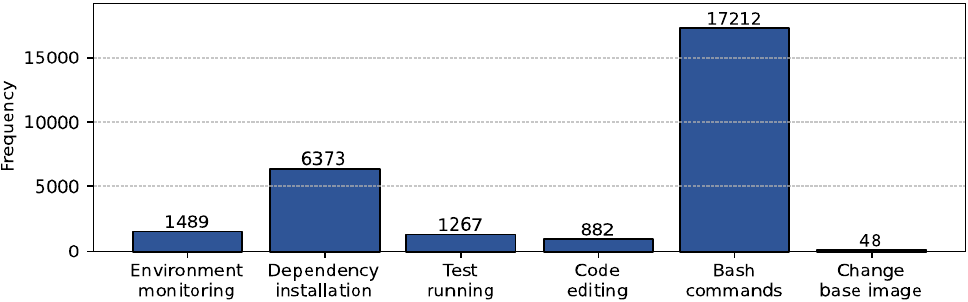}
    \caption{Action usage frequency of the success building.}
\label{figs:tool_frequency}
\end{figure}

As shown in Figure~\ref{figs:tool_frequency}, we analyze the invocation times of various action types in 361 successfully built projects, including the five actions within the internal environment and the action of base image change. Bash commands are the most frequently used action, as they encompass the majority of instructions. Additionally, we observe that the LLM agent tends to call dependency installation quite frequently, averaging about 18 times per building, which means roughly 18 dependencies are installed per building on average. Moreover, the LLM agent calls test running approximately 3.5 times per building on average, which typically helps the agent better identify issues. Among successful buildings, we find 48 instances of changing the base image, accounting for 13.3\%~ of success cases, indicating that the initial selection of the base image is often incorrect and requires subsequent adjustments.

%% file: appendix/synthesizer_example.tex
\subsection{Example}
\begin{figure}[t]
	\centering
	\includegraphics[width=1.0\textwidth]{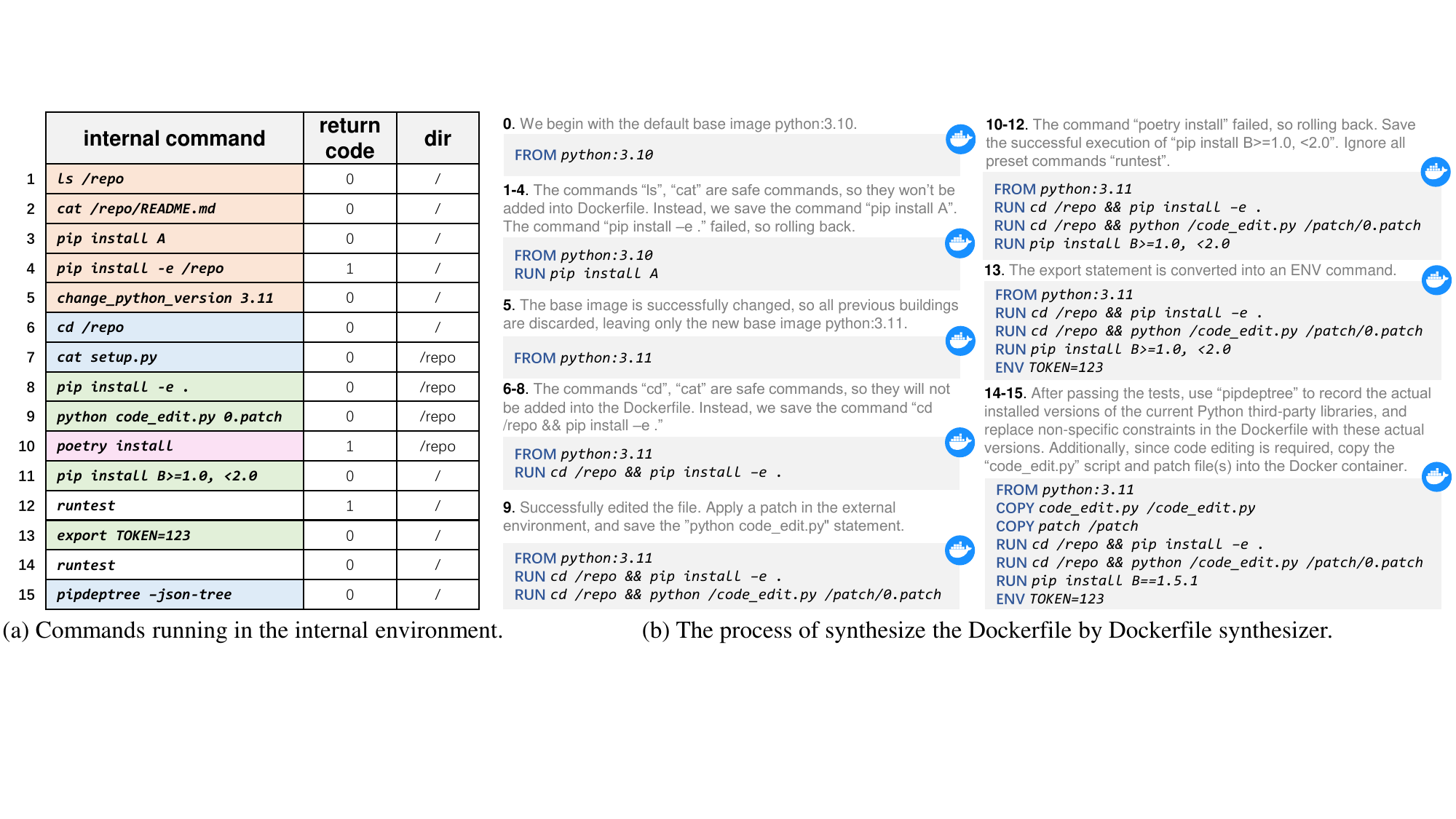}
    
    \caption{An example of Dockerfile synthesizer to transfer the commands into a runnable Dockerfile.}
\label{figs:sequence2dockerfile}
\end{figure}

If the LLM agent successfully runs all tests in the repository in the internal environment, the building process needs to be recorded in a Dockerfile. Figure \ref{figs:sequence2dockerfile} illustrates the Dockerfile synthesizer's operation. Figure~\ref{figs:sequence2dockerfile} (a) lists the internal commands actually executed in the internal environment, their return codes (e.g., \texttt{0} indicates success, otherwise failure), and the current directory (\texttt{dir}). Figure~\ref{figs:sequence2dockerfile} (b) shows the process of generating the Dockerfile.

The principles for the conversion process include the following essential steps: scanning all internal commands sequentially, and the line numbers mentioned below precisely match the line number of Figure \ref{figs:sequence2dockerfile} (a):
\begin{enumerate}[leftmargin=*]
\item By default, use \texttt{python:3.10} as the base image.
\item For commands that run successfully (i.e., have a return code of \texttt{0}), prepend the command with ``\texttt{RUN}'' to form a Dockerfile statement (e.g., line 3). Commands that fail (i.e., have a return code other than \texttt{0}, such as line 4) are rolled back and not included in the Dockerfile. An exception is that some commands typically do not change the current state (such as ``\texttt{cat}'' on line 2). They are not added to the Dockerfile. We show the complete list in Appendix~\ref{safe_commands}.
\item If \texttt{dir} is not the root directory, use \texttt{cd} to change to that directory before running the command, as each Dockerfile statement runs in its own directory session.
\item If a base image change statement is encountered (e.g., line 5), discard all previous buildings and switch to the new base image.
\item If a code editing command is encountered (e.g., line 9), copy the patch and editing script into the Docker container before executing the editing command.
\item If an export statement for adding environment variables is encountered (e.g., line 13), convert it to a persistent \texttt{ENV} statement.
\item After all statements have been scanned, check for dependency installation commands and replace any unspecified versions (e.g., ``\texttt{B>=1.0,<2.0}'' in line 11) with the actual versions downloaded in the Docker container (e.g., ``\texttt{B==1.5.1}'' in the final Dockerfile of Figure~\ref{figs:sequence2dockerfile} (b)).
\end{enumerate}

\subsection{Safe command}
\label{safe_commands}
If \tool~executes the following commands without using ``$>$'' or ``$>>$'' for output redirection, they are regarded to be safe commands that typically do not affect the system. Therefore, rollback is not necessary, and they are not added to the generated Dockerfile.

[``\texttt{cd}'', ``\texttt{ls}'', ``\texttt{cat}'', ``\texttt{echo}'', ``\texttt{pwd}'', ``\texttt{whoami}'', ``\texttt{who}'', ``\texttt{date}'', ``\texttt{cal}'', ``\texttt{df}'', ``\texttt{du}'', ``\texttt{free}'', ``\texttt{uname}'', ``\texttt{uptime}'', ``\texttt{w}'', ``\texttt{ps}'', ``\texttt{pgrep}'', ``\texttt{top}'', ``\texttt{dmesg}'', ``\texttt{tail}'', ``\texttt{head}'', ``\texttt{grep}'', ``\texttt{find}'', ``\texttt{locate}'', ``\texttt{which}'',
``\texttt{file}'', ``\texttt{stat}'', ``\texttt{cmp}'', ``\texttt{diff}'', 
``\texttt{xz}'', ``\texttt{unxz}'', ``\texttt{sort}'', ``\texttt{wc}'', ``\texttt{tr}'', ``\texttt{cut}'', ``\texttt{paste}'', ``\texttt{tee}'', ``\texttt{awk}'', ``\texttt{env}'', ``\texttt{printenv}'',
``\texttt{hostname}'', ``\texttt{ping}'', ``\texttt{traceroute}'', ``\texttt{ssh}'']

%% file: appendix/pollution_commands.tex
\begin{figure}[hbtp]
	\centering
	\includegraphics[width=1.0\textwidth]{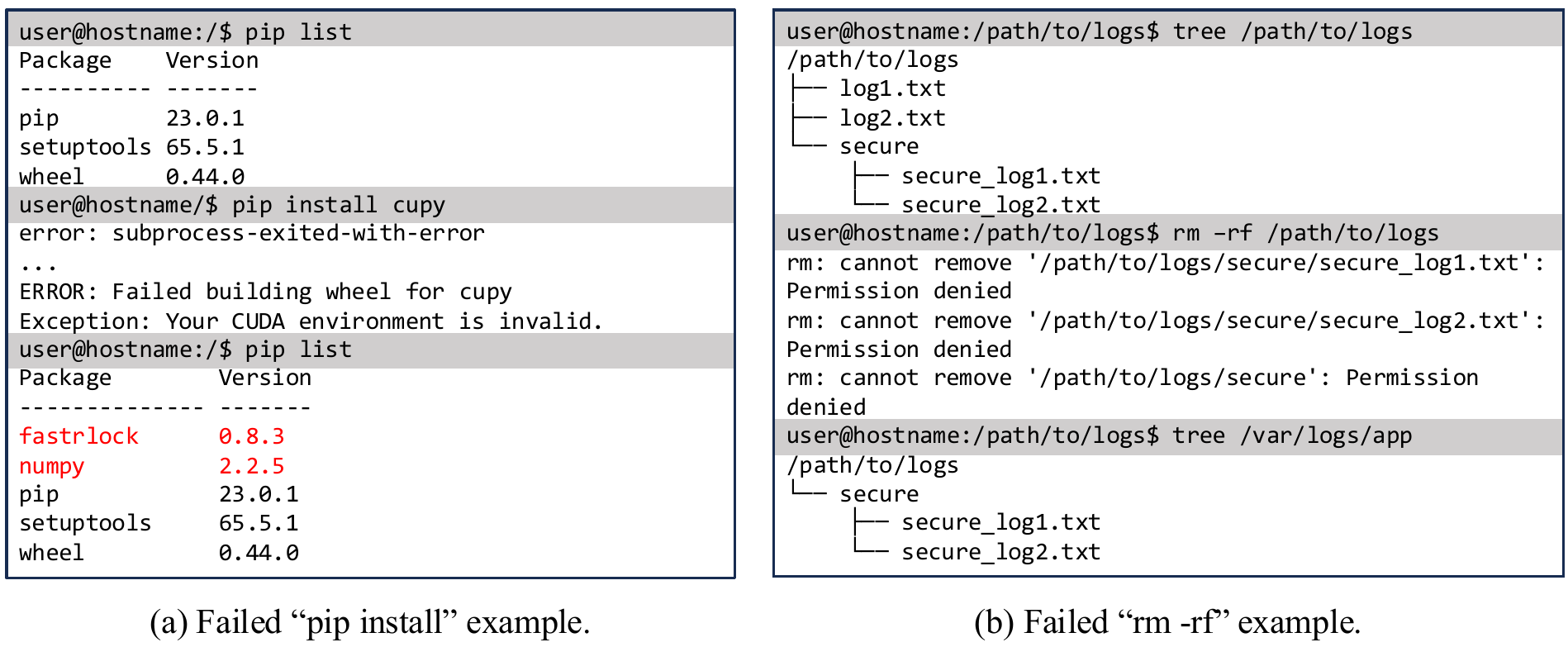}
    \caption{Examples of failed command causing ``pollution'' to the environment.}
\label{figs:pollution_example}
\end{figure}

Figure~\ref{figs:pollution_example} shows two practical examples. As shown
in Figure~\ref{figs:pollution_example} (a), when we run ``\texttt{pip install cupy}'', the command fails to execute because the CUDA environment is invalid. However, it still introduces the ``\texttt{fastrlock}'' and ``\texttt{numpy}'' packages which is not present in the original environment. In Figure~\ref{figs:pollution_example} (b), when we run the ``\texttt{rm -rf /path/to/logs}'' command to delete the directory, some files fail to be deleted due to a lack of permission. However, the files that have already been deleted (e.g., \texttt{log1.txt} and \texttt{log2.txt}) are not restored.
Such ``pollution'' caused by failed commands may make the environment enter the uncertain state.

We also identified 30 Python packages that may introduce ``pollution'' when installing via \texttt{pip}, as listed in Table~\ref{tab:pollution_example}. This process can be easily reproduced in a clean Docker environment (such as the initial environment shown in Figure~\ref{figs:pollution_example} (a)). For example, simply running ``\texttt{pip install adb}'' — even without specifying a version, which defaults to the latest — can cause such ``pollution''. This demonstrates that such scenarios are quite common in real-world development and also underscores the importance of the rollback action in~\tool.

{
\begin{longtable}{p{4cm} p{5cm} c}
\caption{30 examples of "pollution" during pip installation. ``Package\_name'' represents the name of the package that failed to download, "pollution" represents other packages introduced by the failed download, and "pollution" count represents the number of these introduced packages.} \\
\toprule
\rowcolor{gray!20}
\textbf{package\_name} & \textbf{"pollution"} & \textbf{"pollution" count} \\
\midrule
\endfirsthead
\multicolumn{3}{c}{{\bfseries \tablename\ \thetable{} -- continued from previous page}} \\
\toprule
\rowcolor{gray!20}
\textbf{package\_name} & \textbf{"pollution"} & \textbf{"pollution" count} \\
\midrule
\endhead
\midrule \multicolumn{3}{r}{{Continued on next page}} \\ \midrule
\endfoot
\bottomrule
\endlastfoot
zbarlight & pillow & 1 \\
\midrule
mxnet-cu91 & chardet, idna, urllib3 & 3 \\
\midrule
texthero & blis, catalogue, certifi, charset-normalizer, click, \ldots & 41 \\
\midrule
url & six & 1 \\
\midrule
robotframework-ride & certifi, charset-normalizer, idna, numpy, packaging, \ldots & 10 \\
\midrule
adb & libusb1, typing & 2 \\
\midrule
onegov-core & fastcache, mailthon, passlib, polib, pytz, \ldots & 12 \\
\midrule
postal & six & 1 \\
\midrule
changes & requests & 1 \\
\midrule
mxnet-cu75mkl & chardet, idna, urllib3 & 3 \\
\midrule
winpdb & numpy, six & 2 \\
\midrule
slybot & attrs, Automat, certifi, chardet, charset-normalizer, \ldots & 30 \\
\midrule
fbprophet & aiohappeyeyeballs, aiohttp, aiosignal, appdirs, async-timeout, \ldots & 44 \\
\midrule
mxnet-cu75 & chardet, idna, urllib3 & 3 \\
\midrule
libarchive & nose & 1 \\
\midrule
atari-py & numpy, six & 2 \\ \midrule
reppy & cachetools, certifi, charset-normalizer, idna, python-dateutil, \ldots & 8 \\
\midrule
sovrin & leveldb, libnacl, msgpack-python, orderedset, Pympler, \ldots & 15 \\
\midrule
scrapely & numpy, six, w3lib & 3 \\
\midrule
kevinsr & python-version & 1 \\
\midrule
pysurvive & colored, numpy, pillow, psutil, pygtrie, \ldots & 6 \\
\midrule
horovod & cffi, cloudpickle, packaging, psutil, PyYAML, \ldots & 6 \\
\midrule
cupy & fastrlock, numpy & 2 \\
\midrule
face-recognition & face-recognition-models & 1 \\
\midrule
nsot & gunicorn, idna, ipaddress, ipython, itypes, \ldots & 15 \\
\midrule
gooey & colored, numpy, pillow, psutil, pygtrie, \ldots & 6 \\
\midrule
wxpython & numpy, six & 2 \\
\midrule
neuralcoref & annotated-types, blis, boto3, botocore, catalogue, \ldots & 47 \\
\midrule
apache-airflow-backport-providers-apache-hive & apispec, argcomplete, attrs, babel, cached-property, \ldots & 58 \\
\midrule
bcolz & numpy & 1 
\label{tab:pollution_example}
\end{longtable}
}

%% file: appendix/benchmark_statistic.tex
As demonstrated in Table~\ref{tab:loc_staticstic}, we list the statistics on lines of code for repositories in our constructed benchmark.

\begin{longtable}{lccccc}
\caption{Statistics on Lines of Code (LOC) for repositories in our benchmark. The table below illustrates the statistics regarding the lines of code (LOC) of all repositories in the benchmark. It separately counts the number of repositories with LOC greater than 10,000 and greater than 100,000. The last two rows provide reference LOC for well-known industry-level repositories.} \\
\toprule
\cellcolor{gray!20}\textbf{Configuration state} & \cellcolor{gray!20}\textbf{\#LOC > 10,000} & \cellcolor{gray!20}\textbf{\#LOC > 100,000} & \cellcolor{gray!20}\textbf{Medium} & \cellcolor{gray!20}\textbf{Max} \\
\midrule
\endfirsthead
\toprule
\textbf{Configuration state} & \textbf{\#LOC > 10,000} & \textbf{\#LOC > 100,000} & \textbf{Medium} & \textbf{Max} \\
\midrule
\endhead
\bottomrule
\endfoot
\bottomrule
\endlastfoot
Success(361) & 220 (60.9\%) & 27 (7.5\%) & 14,578 & 6,431,084 \\
Fail(59) & 50 (84.7\%) & 12 (30.6\%) & 43,001 & 3,584,542 \\
All(420) & 270 (64.3\%) & 39 (9.3\%) & 15,854 & 6,431,084 \\
\texttt{requests} (reference) & Yes & No & 12,782 & 12,782 \\
\texttt{pandas} (reference) & Yes & Yes & 571,849 & 571,849
\label{tab:loc_staticstic}
\end{longtable}


%% file: appendix/pass_rate_experiment.tex
We randomly sample 50 repositories from our benchmark, successfully run their entire test suites, and manually record the pass rates for each method. Table~\ref{tab:pass_rate} shows the percentage of repositories for which each method achieved a test pass rate exceeding various thresholds (from >=10\% to 100\%).

\begin{longtable}{ccccc}
\caption{Pass rates across tools at varying thresholds. Each cell shows the percentage and the number of repositories in parentheses.} \\
\toprule
\cellcolor{gray!20}\textbf{Pass Rate Threshold} & \cellcolor{gray!20}\textbf{Repo2Run} & \cellcolor{gray!20}\textbf{SWE-agent} & \cellcolor{gray!20}\textbf{LLM generator} & \cellcolor{gray!20}\textbf{pipreqs} \\
\midrule
\endfirsthead
\toprule
\textbf{Pass Rate Threshold} & \textbf{Repo2Run} & \textbf{SWE-agent} & \textbf{LLM generator} & \textbf{pipreqs} \\
\midrule
\endhead
\bottomrule
\endfoot
\bottomrule
\endlastfoot
$\geq$10\% & \textbf{70\% (35)} & 26\% (13) & 32\% (16) & 10\% (5) \\
$\geq$20\% & \textbf{68\% (34)} & 26\% (13) & 32\% (16) & 10\% (5) \\
$\geq$30\% & \textbf{64\% (32)} & 26\% (13) & 32\% (16) & 10\% (5) \\
$\geq$40\% & \textbf{62\% (31)} & 26\% (13) & 32\% (16) & 8\% (4) \\
$\geq$50\% & \textbf{62\% (31)} & 24\% (12) & 32\% (16) & 8\% (4) \\
$\geq$60\% & \textbf{58\% (29)} & 20\% (10) & 30\% (15) & 8\% (4) \\
$\geq$70\% & \textbf{54\% (27)} & 18\% (9) & 28\% (14) & 8\% (4) \\
$\geq$80\% & \textbf{52\% (26)} & 18\% (9) & 26\% (13) & 4\% (2) \\
$\geq$90\% & \textbf{48\% (24)} & 18\% (9) & 26\% (13) & 4\% (2) \\
100\% & \textbf{34\% (17)} & 10\% (5) & 14\% (7) &  4\% (2) 
\label{tab:pass_rate}
\end{longtable}

The results are clear: regardless of the pass rate threshold chosen, \tool consistently and substantially outperforms all baselines. This experiment demonstrates that even when evaluated by the metric you suggested, our method's superiority remains clear.

We conduct this experiment to directly show that \tool excels not just in creating executable environments, but in creating high-quality, correct ones.
We wish to briefly emphasize why we firmly believe that EBSR (Environment Building Success Rate) is an appropriate metric for this task.

Our goal is to automate the crucial first step of software usability: creating an executable environment. Test pass rates, while informative, are a ``noisy'' signal for environment quality because failures can arise from code bugs or test bugs, not just environment building errors. As industry practice (e.g., CI/CD) shows, distinguishing between a build/execution success and a test-pass success is a standard and necessary distinction.

%% file: appendix/manual_experiment.tex
\subsection{Manual Experiment}
\subsubsection{Settings}
To ensure a fair comparison among participants, we conducted training and demonstrated examples before the experiment to ensure everyone understood the procedure. Additionally, to minimize discrepancies in time consumption due to network factors, all participants conducted the experiment in the same network environment.

\subsubsection{Survey}
We selected eight technical staff from internet companies to participate in the experiment and conducted a survey regarding their backgrounds prior to the experiment.
Their development experience ranges from 4 to 11 years, with an average of 7 years in software development and 3.8 years in Python development. Seven participants have experience in complex development projects, while one has experience in multiple small-scale projects.

Regarding environment building, three participants indicated that they spend a significant amount of time building the environment when faced with an unfamiliar code repository; another three stated that, although they spend a long time, it is generally manageable; two participants reported spending minimal time.

In terms of successful environment building in their regular work, three participants mentioned they only fail with extremely complex environments, while five indicated they can build environments for most medium-scale repositories.
As for their confidence in successfully building unfamiliar environments, six participants expressed that they are usually successful, and two said they are sometimes successful.

When it comes to the amount of time they are willing to wait to build an unfamiliar repository, four participants are willing to wait for over 90 minutes, two are willing to wait 40-60 minutes, one is willing to wait 20-40 minutes, and one is only willing to wait 10-20 minutes.

Participants' overall evaluation of building code running environments is diverse: one finds it very troublesome with many issues, four find it somewhat troublesome, two indicate moderate difficulty, and one finds it relatively simple.

Additionally, all participants expressed a high or very high willingness to use a tool that could automatically build the environment for an unfamiliar repository.

\subsubsection{Experiment guideline}
We request all participants to conduct the experiment following the guidelines below:

\textbf{1. Environment Setup}

Build the Docker environment. Verify the installation is successful: If installed correctly, you should be able to use the following command (python:3.10 is just an example; select the base image according to your requirements):

\texttt{docker run -it python:3.10 bash}

\textbf{2. Overall Procedure}

Our objective is to install the given package in a Docker container, build its environment, and be able to run its internal tests. During the process, record the time developers spend building the environment and eventually save the logs using docker logs.

\textbf{2.1 Review the Repository to be Built (Optional)}

Review the GitHub repository that needs to be built.

\textbf{2.2 Determine the Docker Base Image (Generally start with python:3.10)}

Select the Docker base image based on the repository (all repositories use Python as the main language). Common base images include: Note, in this experiment, it is generally sufficient to use the official Python series images. The main concern is the version; if uncertain, you can start with newer versions like 3.10, 3.11, 3.12, or select the recommended Python version based on the repository's README.

Python series (the number after python indicates the version): python:3.10, python:3.12, python:3.9, python:3.6, python:3.9...
PyTorch series: pytorch/pytorch, pytorch/pytorch:1.9.0-cuda10.2-cudnn7-runtime...
Anaconda series: continuumio/anaconda3...
Note: If you find the selected version is incorrect later, you can exit and reselect.

\textbf{2.3 Create and Enter the Container}

Using the determined base image name (e.g., python:3.10), enter the container with the following command:

\texttt{docker run -it --name mytest python:3.10 bash}

Here, mytest is the container name, recorded for log export later. It can be freely named, just keep track of it to avoid losing it later. Note: If any issues arise here, check if Docker is correctly installed and if the image name is valid, and troubleshoot accordingly.

\textbf{2.4 Install Relevant Tools}

APT tool downloads:

\texttt{apt-get update \&\& apt-get install -y curl}

Download pytest:

\texttt{pip install pytest}

\textbf{2.5 Download the Repository}

Select the repository to be built and download its GitHub repository to a location (generally directly in the root directory):

\texttt{git clone https://github.com/\{full\_name\}.git}

\textbf{2.6 Build the Environment}

Now, use your skills to build: First, enter the downloaded file directory, for example:

\texttt{cd wddbfs}

Switch to the specified branch SHA (refer to the corresponding SHA of the repository), for example:

\texttt{git checkout 5c68aa}

Our goal is to successfully run pytest (not necessarily to pass all tests, just to run them). A simple criterion is to successfully run:

\texttt{pytest --collect-only -q}

At this point, you can use your experience and information from the repository documentation and debugging error messages to build. However, there are a few restrictions:

Do not directly edit the test files! (Files starting with test\_ or ending with \_test.py).
Do not directly delete test files!
Editing the original repository files is not recommended.

During this process, you may perform various operations, including but not limited to pip, apt-get, and other tool downloads, as well as searching online or using GPT for debugging help.

Additionally, if there are long download times requiring waiting, you may decide according to your situation whether to leave this running and do other things (just don’t forget about this task).

\textbf{2.7 Completion and Logging}

A task can conclude in two scenarios:

Scenario One: If ``\texttt{pytest --collect-only -q}'' runs without issues, you can then execute pytest. If pytest completes successfully, the task is done.
Scenario Two: If you feel the package is extremely difficult to build, for example exceeding your patience threshold (refer to your usual development habits), you may also terminate.

Once finished, input exit to exit. Make sure to save your output logs with the following command (replace {container\_name} with the container name you recorded earlier, if you forget, you can use docker ps to check):

\texttt{docker logs {container\_name} -t > wddbfs.log}

\textbf{2.8 Fill Out the Form and Record Information}

You need to fill out the form according to your feelings.

Here, Table~\ref{tab:form} presents the form that developers are required to complete after the environment building process, including items such as ``Is it successful?'', ``Final base image used'', ``Reason for failure'', and so on. Besides, Tables~\ref{tab:form_example1} and \ref{tab:form_example2} show two examples filled out by developers.

{
\begin{longtable}{p{3.5cm} p{9.5cm}}
\caption{The form that developers should fill in after the environment building.} \\
\toprule
\rowcolor{gray!20}
\textbf{Question} & \textbf{Description} \\
\midrule
\endfirsthead
\multicolumn{2}{c}%
{{\bfseries \tablename\ \thetable{} -- continued from previous page}} \\
\toprule
\rowcolor{gray!20}
\textbf{Question} & \textbf{Description} \\
\midrule
\endhead
\midrule \multicolumn{2}{r}{{Continued on next page}} \\ \midrule
\endfoot
\bottomrule
\endlastfoot
Is it successful? & Yes or No - whether the task successfully passed ``\texttt{pytest --collect-only -q}'' without errors and eventually ran ``\texttt{pytest}''. \\ \midrule
Final base image used & E.g., python:3.10 \\ \midrule
Reason for failure & Summarize the main reasons for failure (if failed), including:
\begin{itemize}[leftmargin=*,noitemsep,nolistsep]
    \item Long download time
    \item Difficulty handling repository dependencies
    \item Unresolvable bug
    \item Errors in repository tests
    \item Lengthy test durations
    \item Other reasons (please specify)
\end{itemize}\\ 
\midrule
Waiting time & Approximate value, indicating the time spent waiting for dependencies to download. Does not include time spent on decision-making and research. Provide an approximate range:
\begin{itemize}[leftmargin=*,noitemsep,nolistsep]
    \item <3 minutes
    \item 3-5 minutes
    \item 6-10 minutes
    \item 11-20 minutes
    \item 20-40 minutes
    \item 40-60 minutes
    \item 60-90 minutes
    \item 90+ minutes
\end{itemize} \\ \midrule
Longest time-consuming process & Describe the most time-consuming building process, such as:
\begin{itemize}[leftmargin=*,noitemsep,nolistsep]
    \item Downloading a specific dependency
    \item Resolving a specific error
    \item Incorrect Python version selection
\end{itemize} \\ \midrule
Tolerance level & Based on your subjective feeling and development experience. Rate the process (1-5):
\begin{itemize}[leftmargin=*,noitemsep,nolistsep]
    \item 1: Extremely unbearable
    \item 2: Somewhat unbearable, but manageable
    \item 3: Neutral, tolerable
    \item 4: Comfortable, no significant discomfort
    \item 5: Very comfortable, highly satisfactory building experience
\end{itemize} \\ \midrule
Building difficulty & Based on the complexity of the building process (ignoring time spent): Rate the process (1-5):
\begin{itemize}[leftmargin=*,noitemsep,nolistsep]
    \item 1: Very simple, completed with intuition and experience, no reference materials needed
    \item 2: Fairly simple, referred to basic materials (e.g., README), simple overall steps
    \item 3: Moderate difficulty, encountered some issues, but manageable
    \item 4: Difficult, required extensive debugging and building
    \item 5: Very difficult, needed numerous references and encountered unresolved or time-consuming issues
\end{itemize} \\ \midrule
Materials referenced & List the materials used for reference (e.g., README, internal building files, online searches, GPT). If none, mention ``Directory". \\ \midrule
Biggest challenge during the process & Describe the most troublesome aspect of the building process, such as long wait times, unclear error messages, dependency version conflicts, inability to find required software versions, etc.
\label{tab:form}
\end{longtable}
}

\newpage
\textbf{Example 1}:

\begin{longtable}{lp{8cm}}
\caption{A sample form filled by a developer after environment building, illustrating a successful environment building.} \\
\toprule
\cellcolor{gray!20}\textbf{Question} & \cellcolor{gray!20}\textbf{Answer} \\ 
\hline
Is it successful? & Yes \\ 
\hline
Final base image used & python:3.10 \\ 
\hline
Reason for failure & Successful \\ 
\hline
Waiting time & 3-5 minutes \\ 
\hline
Longest time-consuming process & Downloading poetry \\ 
\hline
Tolerance level & 5 \\ 
\hline
Building difficulty & 1 \\ 
\hline
Materials referenced & Directory \\ 
\hline
Biggest challenge during the process & Waiting for building and installation \\ 
\hline
\label{tab:form_example1}
\end{longtable}

\textbf{Example 2}:

\begin{longtable}{lp{8cm}}
\caption{A sample form filled by a developer after environment building, illustrating a failed environment building.} \\
\toprule
\cellcolor{gray!20}\textbf{Question} & \cellcolor{gray!20}\textbf{Answer} \\ 
\hline
Is it successful? & No \\ 
\hline
Final base image used & python:3.11 \\ 
\hline
Reason for failure & Unresolvable bug, provide bug image \\ 
\hline
Waiting time & 40-60 minutes \\ 
\hline
Longest time-consuming process & Resolving a series of FileNotFoundError, ImportError, incorrect Python version selection \\ 
\hline
Tolerance level & 2 \\ 
\hline
Building difficulty & 5 \\ 
\hline
Materials referenced & README, GPT, StackOverflow \\ 
\hline
Biggest challenge during the process & Dependency version conflicts, long wait times \\ 
\bottomrule
\label{tab:form_example2}
\end{longtable}

\subsubsection{Repository assignment}
We randomly assigned each participant four unique code repositories that were successfully built by \tool. Additionally, each participant was assigned two code repositories that \tool~failed to build. To avoid chance occurrences, each failed repository was assigned to two different participants. Below is the list of selections:

\textbf{Successfully built}:

[alexwlchan/safari-webarchiver, ManiMozaffar/aioclock, mixedbread-ai/batched, mobiusml/gemlite, circlemind-ai/fast-graphrag, knowsuchagency/promptic, mbodiai/embodied-agents, modelscope/agentscope, Adibvafa/CodonTransformer, kennethreitz/simplemind, lmstudio-ai/venvstacks, mlecauchois/micrograd-cuda, IST-DASLab/PanzaMail, MetaGLM/zhipuai-sdk-python-v4, openai/mle-bench, RealOrangeOne/django-tasks, basf/MolPipeline, dai-motoki/zoltraak, lucidrains/alphafold3-pytorch, mistralai/mistral-common, BMPixel/moffee, DataformerAI/dataformer, jahwag/ClaudeSync, volfpeter/htmy, Genentech/gReLU, OpenNLPLab/lightning-attention, paradigmxyz/spice, reagento/dishka, arcee-ai/fastmlx, KyanChen/RSMamba, neuralmagic/guidellm, simonw/files-to-prompt]

\textbf{Failed to build}:

[zhuqinfeng1999/Samba, dongxuyue/Open-ReplaceAnything, LazyAGI/LazyLLM, jialuechen/deepfolio, KOSASIH/pi-nexus-autonomous-banking-network, AARG-FAN/Yolo\_for\_Wukong, plinder-org/plinder, expectedparrot/edsl]

\subsubsection{Results}
Based on the times shown in the logs, we calculated that the average building time for each repository is 21.33 minutes. Furthermore, none of the repositories that \tool~failed to build were successfully built manually. Additionally, in the manual experiment, five environments that were successfully built by \tool~were not successfully built, representing 15.6\% of the total successfully built assignments.

The table information and logs are too lengthy, so we place them on \url{https://anonymous.4open.science/r/Repo2Run}.

\subsection{Industry-level repository building validation}
As shown in Table~\ref{tab:industry_evaluation}, we present the evaluation of \tool~ on 59 popular industry-grade repositories. Additionally, Table~\ref{tab:industry_evaluation1} shows the successful configuration of additional industry-grade repositories.

\begin{longtable}{lcccc}
\caption{Evaluation of \tool~on 59 popular industry-grade repositories. The repositories were selected from GitHub Ranking~\cite{githubranking}, filtering out those without tests and with \#LOC < 10,000.} \\
\toprule
\cellcolor{gray!20}\textbf{Configuration state} & \cellcolor{gray!20}\textbf{\#LOC > 10,000} & \cellcolor{gray!20}\textbf{\#LOC > 100,000} & \cellcolor{gray!20}\textbf{Medium} & \cellcolor{gray!20}\textbf{Max} \\
\midrule
\endfirsthead
\toprule
\textbf{Configuration state} & \textbf{\#LOC > 10,000} & \textbf{\#LOC > 100,000} & \textbf{Medium} & \textbf{Max} \\
\midrule
\endhead
\bottomrule
\endfoot
\bottomrule
\endlastfoot
Success(30) & 30 (100\%) & 14 (46.7\%) & 96,157 & 2,389,065 \\
Fail(29) & 29 (100\%) & 24 (82.6\%) & 333,464 & 2,669,973\\
All(59) & 59 (100\%) & 40 (9.3\%) & 124,352 & 2,669,973
\label{tab:industry_evaluation}
\end{longtable}

\begin{longtable}{p{2.6cm} r p{2.6cm} r p{2.6cm} r}
\caption{Successful additional configuration of 30 industry-grade repositories.} \\
\toprule
\rowcolor{gray!20}
\textbf{Repository} & \textbf{\#LOC} & \textbf{Repository} & \textbf{\#LOC} & \textbf{Repository} & \textbf{\#LOC} \\
\midrule
\endfirsthead
\multicolumn{6}{c}%
{{\bfseries \tablename\ \thetable{} -- continued from previous page}} \\
\toprule
\rowcolor{gray!20}
\textbf{Repository} & \textbf{\#LOC} & \textbf{Repository} & \textbf{\#LOC} & \textbf{Repository} & \textbf{\#LOC} \\
\midrule
\endhead
\midrule \multicolumn{6}{r}{{Continued on next page}} \\ \midrule
\endfoot
\bottomrule
\endlastfoot
comfyanonymous/ ComfyUI & 737,543 & OpenBB-finance/OpenBB & 2,389,065 & scrapy/scrapy & 77,278 \\ \midrule
yt-dlp/yt-dlp & 155,662 & keras-team/keras & 204,537 & AUTOMATIC1111/ stable-diffusion-webui & 40,448 \\ \midrule
pallets/flask & 25,190 & geekan/MetaGPT & 82,305 & psf/black & 118,551 \\ \midrule
deepfakes/faceswap & 74,690 & streamlit/streamlit & 201,894 & psf/requests & 12,782 \\ \midrule
lllyasviel/Fooocus & 328,199 & labmlai/annotated\_ deep\_learning \_paper\_implementations & 510,851 & RVC-Boss/GPT-SoVITS & 34,658 \\ \midrule
mingrammer/ diagrams & 11,994 & OpenInterpreter/open-interpreter & 15,818 & hiyouga/LLaMA-Factory & 76,000 \\ \midrule
nvbn/thefuck & 10,267 & pandas-dev/pandas & 571,849 & ytdl-org/youtube-dl & 96,157 \\ \midrule
TheAlgorithms/ Python & 419,917 & fastapi/fastapi & 177,120 & Textualize/rich & 45,744 \\ \midrule
All-Hands-AI/OpenHands & 160,514 & Stability-AI/ stablediffusion & 13,698 & ultralytics/ultralytics & 94,201 \\ \midrule
gradio-app/gradio & 184,722 & QuivrHQ/quivr & 10,264 & freqtrade/freqtrade & 188,889
\label{tab:industry_evaluation1}

\end{longtable}

%% file: appendix/Repo2Run_tools.tex
Showing in Table~\ref{tab:command_list}, we design the following actions for Repo2Run to facilitate its invocation.

\begin{table}[hbtp]
    \centering
    \caption{Command list and their functions.}
    \begin{tabular}{p{6cm}p{7cm}}
        \toprule
        \cellcolor{gray!20}\textbf{Command} & \cellcolor{gray!20}\textbf{Function} \\
        \hline
        \texttt{waitinglist add -p package\_name} \texttt{[-v version\_constraints] -t tool}& Add item into waiting list. If no ``version\_constraints'' are specified, the latest version will be downloaded by default.\\
        \hline
        \texttt{waitinglist addfile file\_path} & Add all entries from a file similar to requirements.txt format to the waiting list. Format should be package\_name [version\_constraints]. \\
        \hline
        \texttt{waitinglist clear} & Clear all items in the waiting list. \\
        \hline
        \texttt{conflictlist solve -v} ``\texttt{[version\_constraints]}'' & Resolve the conflict for the first element in the conflict list, and update the version constraints for the corresponding package\_name and tool to version\_constraints. If no ``version\_constraints'' are specified, the latest version will be downloaded by default. The package\_name and tool in the original waiting list must match one of the elements in the conflictlist. Here, the version\_constraints are specified. \\
        \hline
        \texttt{conflictlist solve -u} & Keep the original version constraint that exists in the waiting list, and discard the other version constraints with the same name and tool in the conflict list. \\
        \hline
        \texttt{conflictlist clear} & Clear all items in the conflict list. \\
        \hline
        \texttt{conflictlist show} & Show all items in the conflict list. \\
        \hline
        \texttt{waitinglist show} & Show all items in the waiting list. \\
        \hline
        \texttt{download} & Download all pending items in the waiting list at once, and the conflict list must be empty before executing. \\
        \hline
        \texttt{runtest} & Check if the built environment is correct using``\texttt{pytest}''. \\
        \hline
        \texttt{poetryruntest} & Check if the built environment is correct in the poetry environment. If you want to run tests in the poetry environment, run it. \\
        \hline
        \texttt{runpipreqs} & Generate \texttt{requirements\_pipreqs.txt} and \texttt{pipreqs\_output.txt} and \texttt{pipreqs\_error.txt}.\\
        \hline
        \texttt{change\_python\_version} \texttt{python\_version} & Switching the Python version in the Docker container will forgo any installations made prior to the switch. The Python version number should be represented directly with numbers and dots, without any quotation marks. \\
        \hline
        \texttt{clear\_configuration} & Reset all the buildings to the initial setting of \texttt{python:3.10}. \\
        \bottomrule
    \end{tabular}
    \label{tab:command_list}
\end{table}

%% file: tables/benchmark.tex
As shown in Table~\ref{tabs:success_status}, we list the success status of all the packages in our constructed benchmark.

\setlength{\tabcolsep}{0.1mm}
\small
\begin{longtable}{p{5cm}ccp{5cm}cc}
\caption{Success status of each package in the benchmark.} \\
\toprule
\cellcolor{gray!20}\textbf{full\_name} & \cellcolor{gray!20}\textbf{sha} & \cellcolor{gray!20}\textbf{success} & \cellcolor{gray!20}\textbf{full\_name} & \cellcolor{gray!20}\textbf{sha} & \cellcolor{gray!20}\textbf{success} \\ \midrule
\endfirsthead

\multicolumn{6}{c}%
{{\bfseries \tablename\ \thetable{} -- continued from previous page}} \\
\toprule
\cellcolor{gray!20}\textbf{full\_name} & \cellcolor{gray!20}\textbf{sha} & \cellcolor{gray!20}\textbf{success} & \cellcolor{gray!20}\textbf{full\_name} & \cellcolor{gray!20}\textbf{sha} & \cellcolor{gray!20}\textbf{success} \\
\midrule
\endhead

\midrule \multicolumn{6}{r}{{Continued on next page}} \\ \midrule
\endfoot

\bottomrule
\endlastfoot
271374667/VideoFusion & 9ba7b8 & Yes & 6abd/horus & c1d093 & Yes \\
a-r-r-o-w/cogvideox-factory & 80d115 & Yes & a-s-g93/neo4j-runway & 16b441 & Yes \\
Aaditya-Prasad/consistency-policy & eed0c4 & No & AARG-FAN/Yolo\_for\_Wukong & 07f61a & No \\
adamobeng/wddbfs & 5c68aa & Yes & Adibvafa/CodonTransformer & 2842ef & Yes \\
AdityaNG/kan-gpt & 0c6e4c & Yes & Admyral-Security/admyral & de332e & Yes \\
AgentOps-AI/AgentStack & ff9c6a & Yes & aidatatools/ollama-benchmark & c6a5fd & Yes \\
AIR-Bench/AIR-Bench & 4b27b8 & Yes & airbytehq/PyAirbyte & 7e65ab & Yes \\
airtai/fastagency & 1ff503 & Yes & Akkudoktor-EOS/EOS & fff685 & Yes \\
alexmolas/microsearch & 632ff2 & Yes & alexwlchan/safari-webarchiver & 0e4974 & Yes \\
AlibabaPAI/llumnix & b319b2 & Yes & All-Hands-AI/OpenHands & 246107 & Yes \\
alvin-r/databonsai & 3f2b7c & Yes & amchii/tg-signer & 926dbb & Yes \\
andrewyng/aisuite & 763996 & Yes & andrewyng/translation-agent & e0fc60 & Yes \\
AnswerDotAI/byaldi & 4583c0 & Yes & AnswerDotAI/rerankers & ecd1f6 & Yes \\
antgroup/agentUniverse & ed8f55 & Yes & apapiu/transformer\_latent\_diffusion & 84a75e & Yes \\
apify/crawlee-python & 267063 & Yes & apple/ToolSandbox & 1a1dc8 & Yes \\
apple/ml-cross-entropy & 1f3ebd & Yes & apple/ml-mdm & 9a5632 & Yes \\
arcee-ai/fastmlx & fd37bc & Yes & argmaxinc/whisperkittools & 03898f & Yes \\
arvindrajan92/DTrOCR & a10aa0 & Yes & astramind-ai/Auralis & c357a1 & Yes \\
atonderski/neuro-ncap & ecdcf2 & Yes & aurelio-labs/semantic-chunkers & 04acc2 & Yes \\
AuvaLab/itext2kg & 941a1d & Yes & awslabs/agent-evaluation & 3df695 & Yes \\
Azure/co-op-translator & a4709e & Yes & Azure-Samples/rag-postgres-openai-python & 61bde7 & Yes \\
bananaml/fructose & 5f24ec & Yes & basf/MolPipeline & 2f9bae & Yes \\
basf/mamba-tabular & af1ea0 & Yes & beatzxbt/mm-toolbox & 728e35 & Yes \\
bellingcat/ShadowFinder & 578d5a & Yes & Benexl/FastAnime & 677f46 & Yes \\
betaacid/FastAPI-Reference-App & 8caeca & Yes & bhavnicksm/chonkie & 990493 & Yes \\
bigcode-project/bigcodebench & aa634d & Yes & Bl3f/yato & 4906b0 & Yes \\
Blealtan/efficient-kan & 7b6ce1 & Yes & block/goose & c497a5 & Yes \\
BMPixel/moffee & 0e643d & Yes & boheumd/MA-LMM & ffe9fa & Yes \\
bytewiz3/TravelGPT & b19b43 & Yes & CausalLearning/ReAct & 7d3665 & No \\
cfahlgren1/observers & d46fdb & Yes & chaidiscovery/chai-lab & b6e7fa & Yes \\
character-ai/prompt-poet & 466432 & Yes & cheahjs/palworld-save-tools & 7dc2c7 & Yes \\
chernyadev/bigym & 72d305 & Yes & chrschy/fact-finder & ca57d1 & Yes \\
circlemind-ai/fast-graphrag & 447511 & Yes & cloudflare/cloudflare-python & 228479 & Yes \\
codefuse-ai/CodeFuse-muAgent & e93924 & No & codeintegrity-ai/mutahunter & f88922 & Yes \\
codematrixer/hmdriver2 & c0d075 & Yes & codeskyblue/tidevice3 & d83c34 & Yes \\
codeskyblue/uiautodev & eb8577 & Yes & Codium-ai/AlphaCodium & eb7577 & Yes \\
Codium-ai/cover-agent & 5c4b88 & Yes & COLA-Laboratory/TransOPT & de8bf3 & No \\
Comfy-Org/comfy-cli & 7711db & Yes & CompEpigen/figeno & 14b904 & Yes \\
computer-agents/agent-studio & d7f6cb & Yes & cosmic-cortex/mlfz & 5bf8d2 & Yes \\
cremebrule/digital-cousins & 49400b & Yes & crewAIInc/crewAI-tools & 873935 & Yes \\
cvg/nerf-on-the-go & 3659e7 & Yes & D-Star-AI/dsRAG & 2d5431 & Yes \\
D4Vinci/Scrapling & 012820 & Yes & DAGWorks-Inc/burr & 79137e & Yes \\
dai-motoki/zoltraak & 4dce44 & Yes & darrenburns/posting & 94feab & Yes \\
DataformerAI/dataformer & 0cf88c & Yes & daxa-ai/pebblo & e67b01 & Yes \\
daya0576/beaverhabits & c01257 & Yes & dbos-inc/dbos-transact-py & d6c6ac & Yes \\
deepsense-ai/db-ally & 26033f & Yes & dendrite-systems/dendrite-python-sdk & 27c9da & Yes \\
denser-org/denser-retriever & 76256e & No & dingo-actual/infini-transformer & 08d0a1 & Yes \\
discord/access & 19e9b1 & Yes & dleemiller/WordLlama & e38d47 & Yes \\
dongxuyue/Open-ReplaceAnything & 83f0ae & No & dottxt-ai/outlines-core & 31ab9f & Yes \\
dottxt-ai/prompts & 3d2689 & Yes & dreadnode/rigging & 82ac80 & Yes \\
droid-dataset/droid\_policy\_learning & 205ff6 & Yes & DS4SD/docling & aee9c0 & Yes \\
dynamiq-ai/dynamiq & 6cca1c & Yes & eakmanrq/sqlframe & 61fda5 & Yes \\
EleutherAI/sae & 0483b5 & Yes & emcf/thepipe & 02e397 & Yes \\
Emerging-AI/ENOVA & b3661d & Yes & EnhancedJax/Bagels & d72d7f & Yes \\
enoch3712/ExtractThinker & 4872a7 & No & epic-open-source/seismometer & b3e812 & Yes \\
epistoteles/TensorHue & 1564fa & Yes & epogrebnyak/justpath & 0aca51 & Yes \\
erezsh/reladiff & d8683b & Yes & etianen/logot & 54e5ef & Yes \\
expectedparrot/edsl & aa7a2d & No & explosion/spacy-layout & 64c6f4 & Yes \\
facebookresearch/audioseal & ea10f5 & Yes & facebookresearch/lightplane & 34fe6c & Yes \\
facebookresearch/spiritlm & 52fb2f & Yes & FalkorDB/GraphRAG-SDK & 250ebe & Yes \\
Fanqi-Lin/Data-Scaling-Laws & bd6941 & No & fastapi/fastapi-cli & bc0840 & Yes \\
fedirz/faster-whisper-server & cbb6c9 & Yes & felafax/felafax & 34a475 & Yes \\
filipstrand/mflux & 627398 & Yes & FlagOpen/FlagGems & ca13b7 & No \\
fmind/cookiecutter-mlops-package & 00fef7 & Yes & foundation-model-stack/fms-fsdp & 408c75 & Yes \\
fpgmaas/cookiecutter-uv & 90de47 & Yes & frdel/agent-zero & 3cefa1 & Yes \\
frostming/tetos & 106ea5 & Yes & Fugaku-LLM/DeepSpeedFugaku & 74753f & No \\
gauge-sh/bridge & 8b3430 & Yes & Genentech/gReLU & efd308 & Yes \\
genomoncology/FuzzTypes & d96243 & Yes & getludic/ludic & a6db96 & Yes \\
getzep/graphiti & 9f3dd5 & Yes & GigaxGames/gigax & c3c209 & Yes \\
gojasper/flash-diffusion & 48e3bc & Yes & gomate-community/TrustRAG & 1334c4 & Yes \\
gomate-community/rageval & 01e258 & Yes & google-deepmind/nanodo & 10aefd & Yes \\
google-deepmind/penzai & fda6cd & Yes & google-deepmind/treescope & dac18b & Yes \\
google-research/timesfm & 02bc2f & Yes & goombalab/hydra & b6b9b7 & Yes \\
gpustack/gpustack & 4f0c67 & Yes & gregpr07/browser-use & 5e545d & Yes \\
groq/groq-python & fa2e13 & Yes & gusye1234/nano-graphrag & 18fa3a & Yes \\
hailo-ai/hailo-rpi5-examples & 82cfc8 & Yes & Haiyang-W/GiT & ef2b64 & No \\
HanaokaYuzu/Gemini-API & e8a2d2 & Yes & HATTER-LONG/Verbiverse & 82f988 & Yes \\
hinthornw/trustcall & eaaaad & Yes & HKUDS/HiGPT & 2b0793 & No \\
HKUDS/UrbanGPT & be3515 & No & hngprojects/ \quad hng\_boilerplate\_python\_fastapi\_web & bc9740 & Yes \\
hpcaitech/Open-Sora & 38de63 & Yes & hpcaitech/SwiftInfer & 239fd3 & No \\
hrnoh24/stream-vc & faa629 & Yes & huchenlei/ComfyUI\_omost & 7ef00d & Yes \\
huggingface/lerobot & 4c41f6 & Yes & huggingface/lighteval & 6ad727 & Yes \\
HZAI-ZJNU/Mamba-YOLO & ea97fc & No & IAAR-Shanghai/Grimoire & 3fe89d & Yes \\
ib-api-reloaded/ib\_async & 38cf54 & Yes & IBM/fastfit & 396611 & Yes \\
IBM/terratorch & 16e5af & Yes & IEIT-Yuan/Yuan2.0-M32 & b403a2 & No \\
igorbenav/SQLModel-boilerplate & 2ead04 & Yes & igorbenav/fastcrud & dc831b & Yes \\
igrek51/wat & 0d6079 & Yes & illuin-tech/colpali & e45c4c & Yes \\
illuin-tech/vidore-benchmark & 469665 & Yes & Indoxer/LKAN & 16c48e & No \\
Infini-AI-Lab/Sequoia & 688079 & No & instanseg/instanseg & 0df8b2 & Yes \\
instructlab/instructlab & c978b2 & Yes & Integuru-AI/Integuru & 928e82 & Yes \\
InternLM/InternEvo & 5ad2eb & Yes & invariantlabs-ai/invariant & 81547a & Yes \\
IST-DASLab/PanzaMail & b1807c & Yes & iterative/datachain & b67d59 & Yes \\
IvanDrokin/torch-conv-kan & 7a0e83 & Yes & jahwag/ClaudeSync & 000633 & Yes \\
jgravelle/pocketgroq & e995c4 & Yes & jhj0517/AdvancedLivePortrait-WebUI & a7975c & Yes \\
jialuechen/deepfolio & 15d247 & No & jina-ai/late-chunking & db558c & Yes \\
jlowin/fastmcp & baa300 & Yes & jmschrei/tangermeme & a96897 & Yes \\
jonbarron/camp\_zipnerf & 8e6d57 & Yes & JosephBARBIERDARNAL/pypalettes & 826930 & Yes \\
JoshuaC215/agent-service-toolkit & c72f48 & Yes & jshuadvd/LongRoPE & eb9aba & Yes \\
jxnl/n-levels-of-rag & 2ce110 & Yes & karpathy/minbpe & 1acefe & Yes \\
kennethreitz/simplemind & 39b5a5 & Yes & kevinzakka/mink & cf1a30 & Yes \\
knowsuchagency/promptic & a1930c & Yes & koaning/uvtrick & 2d7f27 & Yes \\
kohjingyu/search-agents & 7c35ac & No & KOSASIH/pi-nexus-autonomous-banking-network & 7fcff4 & No \\
kotaro-kinoshita/yomitoku & 71c85b & Yes & KruxAI/ragbuilder & db3d3d & No \\
kujirahand/tkeasygui-python & b1f293 & No & KyanChen/RSMamba & 3fa305 & Yes \\
kyegomez/MultiModalMamba & 58db40 & No & landing-ai/vision-agent & 63eab8 & Yes \\
langchain-ai/langchain-postgres & 064e5b & Yes & lavague-ai/LaVague & b3557f & Yes \\
LazyAGI/LazyLLM & e0dd38 & No & lenML/Speech-AI-Forge & 0b31b2 & Yes \\
leopiney/neuralnoise & c0313f & Yes & lichao-sun/Mora & 7a030e & No \\
Lightning-AI/litdata & 0a97de & Yes & lightonai/pylate & 8de184 & No \\
LilianHollard/LeYOLO & 872841 & Yes & line/lighthouse & ba9da7 & Yes \\
LlmKira/fast-langdetect & 5728ba & Yes & LMCache/LMCache & 7d3443 & Yes \\
lmstudio-ai/mlx-engine & daeb7a & Yes & lmstudio-ai/venvstacks & 235ce3 & Yes \\
lucasdelimanogueira/PyNorch & ed391e & Yes & LucasFaudman/apkscan & 3b3e62 & Yes \\
lucidrains/alphafold3-pytorch & 49f7c9 & Yes & lucidrains/infini-transformer-pytorch & 5774bb & Yes \\
lucidrains/pi-zero-pytorch & 8ad66f & Yes & lucidrains/titok-pytorch & 2f9525 & Yes \\
lucidrains/transfusion-pytorch & 16f73e & Yes & MadcowD/ell & 36ca5e & Yes \\
ManiMozaffar/aioclock & 3d196b & Yes & MaoXiaoYuZ/Long-Novel-GPT & e952ac & No \\
Marker-Inc-Korea/AutoRAG & aa0bfb & Yes & martius-lab/hitchhiking-rotations & 45b49f & Yes \\
mbodiai/embodied-agents & 8715f6 & Yes & McGill-NLP/weblinx & 6f2014 & Yes \\
McGill-NLP/webllama & 696a7c & Yes & Menghuan1918/pdfdeal & e08199 & Yes \\
meta-llama/llama-stack-apps & f14a73 & Yes & MetaGLM/zhipuai-sdk-python-v4 & 7ff4de & Yes \\
metavoiceio/metavoice-src & de3fa2 & Yes & microsoft/MInference & 7a3e5a & No \\
microsoft/TinyTroupe & 9b8d4e & Yes & microsoft/Trace & 826cf5 & Yes \\
microsoft/aurora & 8b1165 & Yes & microsoft/graphrag & de1252 & Yes \\
microsoft/semantic-link-labs & 8e37ef & Yes & mikekelly/AgentK & e9ec89 & Yes \\
Mindinventory/MindSQL & 3d0ff0 & Yes & MinishLab/model2vec & 4e3fba & Yes \\
miquido/draive & 270f0c & Yes & mistralai/mistral-common & 5cac5e & Yes \\
mistralai/mistral-finetune & 656df1 & Yes & mixedbread-ai/baguetter & a6e915 & Yes \\
mixedbread-ai/batched & 1a1797 & Yes & mkjt2/lockbox & 58430d & Yes \\
mlecauchois/micrograd-cuda & ab1ca0 & Yes & MLT-OSS/open-assistant-api & 44eeaf & Yes \\
mlx-graphs/mlx-graphs & 4619d9 & No & mobiusml/gemlite & 5ebcca & Yes \\
ModelCloud/GPTQModel & a5aefc & No & modelcontextprotocol/python-sdk & aaf32b & Yes \\
modelscope/MemoryScope & 330b76 & Yes & modelscope/agentscope & ceaf89 & Yes \\
modern-python/that-depends & 65e656 & Yes & muchdogesec/history4feed & 614182 & Yes \\
muditbhargava66/PyxLSTM & f3c9bb & Yes & narwhals-dev/narwhals & a2088f & Yes \\
nasa-jpl/rosa & 5471dc & Yes & neo4j/neo4j-graphrag-python & 0ac06b & Yes \\
neuralmagic/AutoFP8 & e94461 & Yes & neuralmagic/guidellm & ecf298 & Yes \\
NewT123-WM/tnlearn & 50ee75 & Yes & NexaAI/nexa-sdk & 33f6ba & No \\
nicobrenner/commandjobs & 4c7264 & Yes & Nike-Inc/koheesio & 9bd29e & Yes \\
nlmatics/nlm-ingestor & c72542 & Yes & NLPJCL/RAG-Retrieval & d73057 & No \\
nomic-ai/contrastors & 496a05 & No & NousResearch/finetuning-subnet & e2f5eb & Yes \\
NUS-HPC-AI-Lab/VideoSys & 6c92ae & No & NVIDIA/Megatron-Energon & 28aa3b & Yes \\
NVIDIA/NeMo-Skills & 5591f3 & Yes & NVIDIA/kvpress & 715f8a & Yes \\
NVIDIA/logits-processor-zoo & db179a & Yes & NVIDIA/nv-ingest & eec9fa & No \\
NVlabs/Sana & 41dcbe & Yes & NVlabs/VILA & ec7fb2 & No \\
NVlabs/nvTorchCam & cc27be & Yes & ogkalu2/comic-translate & 1933d1 & Yes \\
Open-Wine-Components/umu-launcher & b0c0d4 & Yes & openai/mle-bench & 51ec2b & Yes \\
openai/swarm & 9db581 & Yes & opendatalab/MinerU & 391a99 & Yes \\
openfoundry-ai/model\_manager & 34f9ff & Yes & opengeos/HyperCoast & c1604c & Yes \\
OpenInterpreter/aifs & 3f74c6 & Yes & OpenNLPLab/lightning-attention & d74395 & Yes \\
openpsi-project/ReaLHF & 62d9cd & Yes & openrecall/openrecall & 225a27 & Yes \\
OpenSPG/KAG & 68b2c6 & No & orbital-materials/orb-models & 251573 & Yes \\
outspeed-ai/outspeed & 049b40 & Yes & OwlAIProject/Owl & 919226 & Yes \\
PacktPublishing/LLM-Engineers-Handbook & ec6717 & Yes & paradigmxyz/spice & e962a9 & Yes \\
patched-codes/patchwork & c9b02b & Yes & paulrobello/parllama & 421238 & Yes \\
PeiJieSun/NESCL & 365d20 & Yes & plinder-org/plinder & 9658cc & No \\
pomonam/kronfluence & 884255 & Yes & PrefectHQ/ControlFlow & f259fa & Yes \\
PrimeIntellect-ai/OpenDiloco & 71f5c2 & Yes & PrimeIntellect-ai/prime & a974cf & Yes \\
princeton-nlp/SWE-agent & 8b3571 & Yes & proger/accelerated-scan & db7145 & Yes \\
pydantic/logfire & 3d7924 & Yes & pymupdf/RAG & b25718 & No \\
pytorch-labs/LeanRL & a416e6 & Yes & raphaelmansuy/code2prompt & 3b377b & No \\
RapidAI/RapidDoc & 5e5fef & Yes & RapidAI/RapidLayout & 8e9677 & Yes \\
reagento/dishka & 2ed985 & Yes & real-stanford/universal \qquad \qquad \qquad \_manipulation\_interface & 298776 & No \\
Realiserad/fish-ai & f32c7f & Yes & RealOrangeOne/django-tasks & e6d26c & Yes \\
reidjs/text-scheduler & 8bb7d6 & Yes & reka-ai/reka-vibe-eval & 93ecd9 & Yes \\
remigenet/TKAN & 8a1de0 & Yes & rio-labs/rio & eda40a & Yes \\
robocasa/robocasa & 27f992 & Yes & RobotecAI/rai & d15910 & No \\
robusta-dev/holmesgpt & c4743a & Yes & royreznik/rexi & f1dca8 & Yes \\
run-llama/llama\_deploy & 47efff & Yes & run-llama/llama\_extract & 89438f & Yes \\
run-llama/llama\_parse & f78186 & Yes & SamKhoze/ComfyUI-DeepFuze & edd7fe & No \\
ScrapeGraphAI/Scrapegraph-ai & bae92b & Yes & seanchatmangpt/dspygen & 69f305 & No \\
serverless-ca/terraform-aws-ca & 2da837 & No & ServerlessLLM/ServerlessLLM & 8f1e6b & Yes \\
ServiceNow/BrowserGym & 12aa5e & Yes & ServiceNow/TapeAgents & 3eca5c & Yes \\
ServiceNow/WorkArena & 0ab9cb & No & ShaShekhar/aaiela & 4e8d6a & No \\
shawntan/scattermoe & 63b76a & Yes & ShoggothAI/motleycrew & 19837e & Yes \\
showlab/computer\_use\_ootb & 419d9d & Yes & shun-liang/yt2doc & 201ec2 & Yes \\
siliconflow/BizyAir & cdb3bb & Yes & simonw/files-to-prompt & f9a4d8 & Yes \\
simonw/llm-claude-3 & c62bf2 & Yes & simonw/llm-cmd & 74fb98 & Yes \\
simonw/llm-jq & beaada & Yes & simular-ai/Agent-S & ca83be & Yes \\
sirocco-ventures/raggenie & 99dfe5 & Yes & souzatharsis/podcastfy & 804a61 & Yes \\
SpecterOps/cred1py & 432f91 & Yes & StacklokLabs/promptwright & 42f69b & Yes \\
steinathan/reelsmaker & 75369c & Yes & stephengpope/no-code-architects-toolkit & ffc1a8 & Yes \\
StonyBrookNLP/appworld & bc9c47 & Yes & Storia-AI/sage & f47fa4 & No \\
superlinear-ai/raglite & b02c5a & Yes & swarmzero/swarmzero & 6fcd7a & Yes \\
tahnok/colmi\_r02\_client & 84d3a6 & Yes & taketwo/llm-ollama & dd616e & Yes \\
taobojlen/django-zeal & 232987 & Yes & TencentARC/BrushNet & 101dc3 & No \\
TheAiSingularity/graphrag-local-ollama & bcb98d & Yes & thousandbrainsproject/tbp.monty & a39a26 & Yes \\
THU-MIG/yolov10 & 6fbaf4 & Yes & thu-nics/MoA & da034c & No \\
tjmlabs/AgentRun & 1997dd & Yes & tobiasfshr/map4d & 0b8bcd & Yes \\
Toloka/dbt-af & e7f436 & Yes & TorchJD/torchjd & 1eaafe & Yes \\
tox-dev/tox-uv & d7405a & Yes & TuragaLab/flybody & 2e1088 & Yes \\
turbo-llm/turbo-alignment & 009574 & Yes & TY-Cheng/torchvinecopulib & c3a477 & Yes \\
ucbepic/docetl & 00a761 & Yes & ultrasev/llmproxy & 1a1100 & Yes \\
uname-n/deltabase & 5eafb9 & Yes & unifyai/unify & ea2088 & No \\
Vashkatsi/deply & 6d6875 & Yes & VideoVerses/VideoTuna & ffc6df & Yes \\
vintasoftware/django-ai-assistant & 5b26c7 & Yes & virattt/financial-datasets & 985664 & Yes \\
vllm-project/llm-compressor & 606aab & Yes & volcengine/verl & ed2eaf & Yes \\
volfpeter/fasthx & e850b9 & Yes & volfpeter/htmy & 0322a3 & Yes \\
vysakh0/dravid & 25b03b & Yes & warmshao/FasterLivePortrait & 6aa810 & Yes \\
weareprestatech/hotpdf & 55ab97 & Yes & web-arena-x/visualwebarena & 89f5af & No \\
Weixiang-Sun/Bora & c08bb6 & Yes & whyhow-ai/knowledge-graph-studio & c41043 & Yes \\
whyhow-ai/rule-based-retrieval & 91701f & Yes & whyhow-ai/whyhow & 63a3c6 & Yes \\
WongKinYiu/YOLO & b96c8e & Yes & WU-CVGL/BAD-Gaussians & bdd8b3 & Yes \\
wy-z/container-vm & 07d402 & Yes & xdit-project/xDiT & a7bd51 & Yes \\
xhluca/bm25s & c4fef2 & Yes & yihong0618/klingCreator & e567c6 & Yes \\
yihong1120/Construction-Hazard-Detection & f5e1ca & Yes & yinjunbo/IS-Fusion & 86c882 & No \\
YUCHEN005/GenTranslate & 62e59d & No & YUCHEN005/RobustGER & ad4e37 & No \\
yurujaja/pangaea-bench & e1d12e & Yes & ZenGuard-AI/fast-llm-security-guardrails & 6a867c & Yes \\
zeroasiccorp/logik & ca4bb1 & Yes & zhuqinfeng1999/Samba & 229687 & No \\
zipnn/zipnn & 007319 & Yes & zou-group/textgrad & b2dc68 & Yes
\label{tabs:success_status}

\end{longtable}

%% file: appendix/baseline_settings.tex
\subsection{Pipreqs settings}
\label{pipreqs_settings}
\begin{figure}[h]
	\centering
	\includegraphics[width=1.0\textwidth]{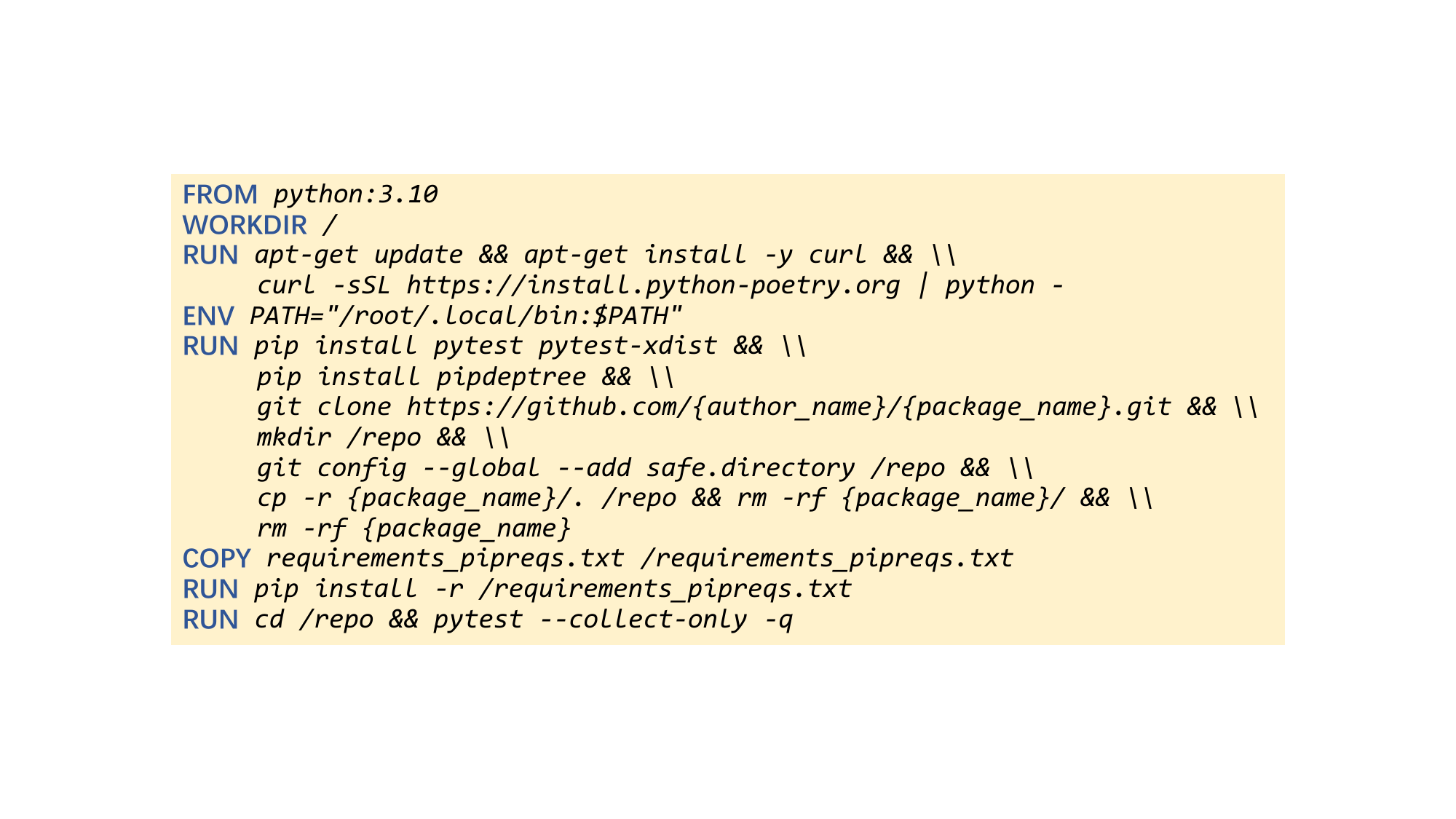}
    
    \caption{Dockerfile prompt using the ``\texttt{requirements\_pipreqs.txt}'' generated by pipreqs.}
\label{figs:dockerfile_prompt}
\end{figure}

Figure~\ref{figs:dockerfile_prompt} shows the template of a Dockerfile generated using ``\texttt{requirements\_pipreqs.txt}'' created by pipreqs.

\subsection{SWE-agent settings}
\label{swe_agent_setting}
Below is the settings of SWE-agent for building the environment: 
\begin{lstlisting}[basicstyle=\ttfamily\small, breaklines=true]
agent:
  templates:
    system_template: |-
      SETTING: You are an expert skilled in environment configuration, operating directly in the command line with a specialized interface.
      The specialized interface consists of {{WINDOW}} lines of context at a time. In addition to typical bash commands, you can also use the following commands to help you with environment setup.
      COMMANDS:
      {{command_docs}}
      Please note that THE EDIT COMMAND REQUIRES PROPER INDENTATION. Ensure correct indentation when modifying code.
      RESPONSE FORMAT:
      Your shell prompt is formatted as follows:
      (Current task: <task name>) <cwd> $
      You need to format your output using two fields: discussion and command.
      Example:
      DISCUSSION
      I'll start by using ls to see the files in the root directory.
      """
      ls /repo
      """
      You should only include a SINGLE command in the command section and then wait for a response. Everything in the DISCUSSION section will be saved for future reference.
      You can use any other bash commands and the special commands listed above, but interactive session commands (e.g., python, vim) are not supported.
    instance_template: |-
      We're currently setting up the environment for the following task. Here are the details:
      TASK:
      {{task_details}}
      INSTRUCTIONS:
      Now, you'll carry out this task on your own. Your terminal session has begun in the repository's root directory. Use the provided commands and any bash commands you need. Edit and check files as needed.
      The goal is to generate a Dockerfile that can successfully build and run the tests in the repository using the command "pytest --collect-only -q".
      NOTE:
      1. The repository is cloned into /repo.
      2. The Dockerfile should start with the following lines (if the base image is python:3.10 and repository is adamobeng/wddbfs):
      """
      FROM python:3.10
      RUN pip install pytest
      RUN git clone https://github.com/adamobeng/wddbfs.git
      RUN mkdir /repo
      RUN git config --global --add safe.directory /repo
      RUN cp -r /wddbfs/. /repo && rm -rf /wddbfs/
      RUN rm -rf /wddbfs
      """
      Your task includes:
      0. **Generate Dockerfile**: Create a file named "Dockerfile" in the root path (e.g, the absolute path is /Dockerfile).
      1. **Read Directory Structure**: Check the folder structure in the root directory.
      2. **Check the Configuration Files**: Inspect files like "requirements.txt", "setup.py", "setup.cfg", "Pipfile*", etc.
      3. **Determine Package Dependencies**: Handle dependencies and manage conflicting dependency versions.
      4. **Testing**: Ensure "pytest /repo --collect-only -q" runs without errors.
      5. **Generate Dockerfile**: Write necessary installation or setup steps determined from the inspection. If you finish run testing successfully, you should modify the Dockerfile in /Dockerfile.
      IMPORTANT TIPS:
      * Check the directory and files carefully.
      * Make sure Dockerfile commands are correct.
      * Use proper Dockerfile syntax and indentation.
      * Test the final Dockerfile by running "pytest /repo --collect-only -q".
      (Current task: Generate Dockerfile)
      (Current directory: {{working_dir}})
      bash-$
    next_step_template: |-
      {{observation}}
      (Current task: Generate Dockerfile)
      (Current directory: {{working_dir}})
      bash-$
    next_step_no_output_template: |-
      Your command ran successfully and did not produce any output.
      (Current task: Generate Dockerfile)
      (Current directory: {{working_dir}})
      bash-$
  tools:
    env_variables:
      WINDOW: 100
      OVERLAP: 2
    bundles:
      - path: tools/registry
      - path: tools/defaults
      - path: tools/search
      - path: tools/edit_linting
      - path: tools/submit
      - path: tools/env_setup
    parse_function:
      type: thought_action
  history_processors:
    - type: last_n_observations
      n: 5
\end{lstlisting}

%% file: appendix/failure_case.tex
\begin{figure}[h]
	\centering
	\includegraphics[width=0.7\textwidth]{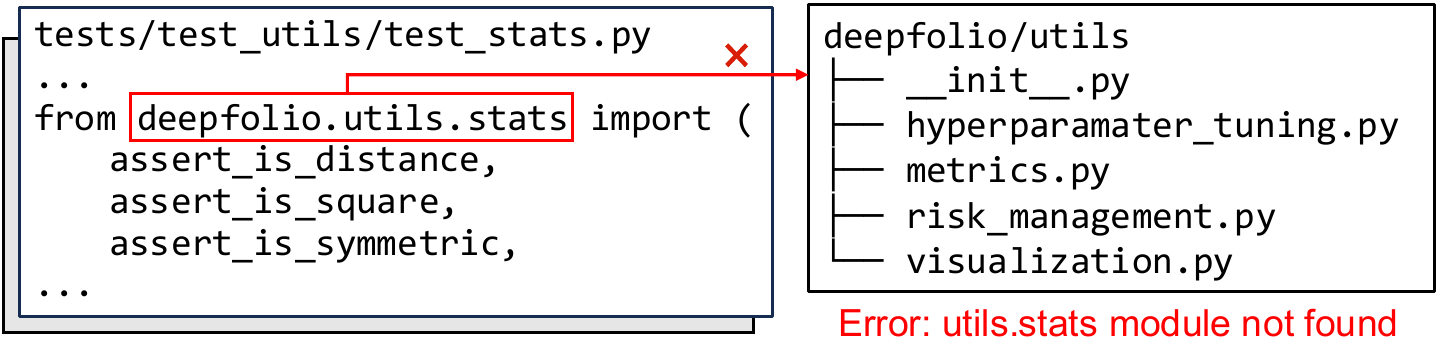}
    \caption{An example of ``\texttt{module not found}'' error due to the absence of updating the unit tests in the repository.}
\label{figs:case_study}
\end{figure}

Figure \ref{figs:case_study}~illustrates a ``module not found'' error when building the repository ``\texttt{jialuechen/deepfolio}''. In this case, the issue arises from the absence of updating the unit tests. The test file ``\texttt{test\_stats.py}'' attempts to import modules from ``\texttt{deepfolio.stats}'', but ``\texttt{stats}'' does not exist in ``\texttt{deepfolio}''. Consequently, no matter how the LLM agent operates, it cannot directly run this test. This highlights the importance for developers to continuously update existing unit tests as the code repository evolves.

